\documentstyle[emulateapj,danonecolfloat,epsfig]{article}

\def\NoApjSectionMarkInTitle#1{#1.\ }
\newcommand\lsim{\mathrel{\rlap{\lower4pt\hbox{\hskip1pt$\sim$}}
        \raise1pt\hbox{$<$}}}
\newcommand\gsim{\mathrel{\rlap{\lower4pt\hbox{\hskip1pt$\sim$}}
        \raise1pt\hbox{$>$}}}

\begin{document}

\twocolumn[


\journalid{337}{15 January 1989}
\articleid{11}{14}

\submitted{\today. To be submitted to ApJ. LA-UR-04-0072}

\title{Tightening Constraints from the Lyman Alpha Forest with the Flux
Probability Distribution Function}

\author{Adam Lidz$^{1,2}$, Katrin Heitmann$^{3}$, Lam Hui$^{1,4}$,
Salman Habib$^{5}$, Michael Rauch$^{6}$, and Wallace
L.~W. Sargent$^{7}$}   
\affil{$^1$ Department of Physics, Columbia University, New York, NY 10027\\
$^2$ Institute for Theory and Computation, Harvard-Smithsonian Center
for Astrophysics, Cambridge, MA 02138\\ 
$^3$ ISR-1, ISR Division, The University of California, Los
Alamos National Laboratory, Los Alamos, NM 87545\\ 
$^4$ Theoretical Astrophysics, Fermi National Accelerator Laboratory,
Batavia, IL 60510\\
$^5$ T-8, Theoretical Division, The University of California, Los
Alamos National Laboratory, Los Alamos, NM 87545\\  
$^6$ Carnegie Observatories, 813 Santa Barbara Street, Pasadena, CA 91101\\
$^7$ Department of Astronomy, California Institute of Technology,
Pasadena, CA 91125\\  
Electronic mail: {\tt alidz@cfa.harvard.edu, heitmann@lanl.gov,
lhui@astro.columbia.edu, habib@lanl.gov, mr@ociw.edu,
wws@phobos.caltech.edu}}

\begin{abstract} 

The analysis of the Lyman-alpha (Ly$\alpha$) forest of absorption
lines in quasar spectra has emerged as a potentially powerful
technique to constrain the linear matter power spectrum. In most
previous work, the amplitude of the ionizing background was fixed by
calibrating simulations to match the observed mean transmitted flux in
the Ly$\alpha$ forest. This procedure is undesirable in principle as
it requires the estimation of the unabsorbed quasar continuum level, a
difficult undertaking subject to various sources of systematic error
and bias. We suggest an alternative approach based on measuring the
one-point probability distribution function (pdf) of the fluctuations in the
flux about the mean, relative to the mean, i.e. the pdf of $\delta_f =
(f - \langle f \rangle)/\langle f \rangle$. This statistic, while
sensitive to the amplitude of the ionizing background, has the virtue
that its measurement does not require an estimate of the unabsorbed
continuum level. We present a measurement of the pdf of $\delta_f$
from seven Keck HIRES spectra, spanning a redshift range of $z=2.2 -
4.4$. To illustrate that our method is useful, we compare our
measurements of the pdf of $\delta_f$, and measurements of the flux
power spectrum from Croft et al. (2002) at $z = 2.72$, with
cosmological simulations. From this comparison, we obtain constraints
on the mean transmission in the Ly$\alpha$ forest, the slope of the
temperature-density relation, as well as the amplitude and slope of
the mass power spectrum. Our methodology will be useful for obtaining
more precise constraints with larger data samples from the Sloan
Digital Sky Survey (SDSS).

\end{abstract}

\keywords{cosmology: theory -- intergalactic medium -- large-scale
structure of universe; quasars -- absorption lines}
]

\section{Introduction}
\label{intro}

Tremendous progress has been made recently in using the Ly$\alpha$
forest as a cosmological probe.  In the emerging theoretical picture
of the Ly$\alpha$ forest, most of the structure in the forest is
attributed to the gravitational instability.  The physics of the
absorbing gas is simple; it is just that of gas in photoionization
equilibrium with a spatially-homogeneous radiation field.  On large
scales the hydrogen gas distribution follows the dark matter
distribution, and on small scales it is Jeans pressure-smoothed (see
e.g., Cen et al.  1994, Zhang et al.  1995, Hernquist et al.  1996,
Miralda-Escud\'e et al.  1996, Muecket et al.  1996, Bi \& Davidsen
1997, Bond \& Wadsley 1997, Hui et al.  1997, Croft et al.  1998,
Theuns et al.  1999, Bryan et al.  1999, Nusser \& Haehnelt 1999).  In
this theoretical picture, each quasar spectrum essentially provides a
one-dimensional map of the density field in the intergalactic medium
(IGM) at $z \sim 3$, implying constraints on the amplitude and slope
of the linear matter power spectrum at $z \sim 3$ and scales of $k
\sim 0.1 - 5 h$ Mpc$^{-1}$ (see e.g., Croft et al.  1998, McDonald et
al.  2000, Croft et al.  2002).

This is a particularly valuable probe since one can infer the {\it
linear} power spectrum of density fluctuations on scales smaller than
those examined by current Cosmic Microwave Background (CMB)
experiments and galaxy surveys (Croft et al.  1998).  The measurements
of the Ly$\alpha$ forest flux power spectrum thus provide an important
complement to other probes of large-scale structure, when constraints
on small-scale clustering from the Ly$\alpha$ forest are combined with
constraints on large scales from CMB experiments.  The leverage from
the large range of scales probed allows for constraints on neutrino
mass (Croft, Hu \& Dav\'e 1999, McDonald 2004a, 2004b, Seljak et
al. 2004), on warm dark matter (Narayanan et al.  2000, Viel et
al. 2005), and on the possibility that the primordial power spectrum
deviates from a pure power law (Spergel et al.  2003, Seljak,
McDonald, \& Makarov 2003, hereafter SMM03, McDonald 2004a, 2004b,
Seljak et al. 2004, Viel et al. 2004c).

However, to obtain reliable constraints on the linear matter power
spectrum from the Ly$\alpha$ forest, a range of sources of systematic
error need to be addressed in more detail.  One such source of error
is that most previous analyses relied on estimating the unabsorbed
quasar continuum level, i.e., the quasar flux in the absence of
absorption from the Ly$\alpha$ forest.  In particular, measurements of
the mean transmitted flux through the Ly$\alpha$ forest were used to
calibrate the amplitude of the photo-ionizing background, a crucial
input parameter in the theoretical modeling of the Ly$\alpha$
forest. This amplitude can only be constrained from observations in an
indirect manner, and it is not feasible to predict this quantity from
first principles with anywhere close to the required accuracy.
Unfortunately, as SMM03 emphasize, it is difficult to robustly measure
the mean transmitted flux in the Ly$\alpha$ forest and hence to
constrain the amplitude of the ionizing background with high accuracy.

In order to elaborate further, we briefly describe the measurements of
the mean transmitted flux and their difficulties.  Two main approaches
have been used to estimating the unabsorbed continuum level (see e.g.,
SMM03).  One approach is to extrapolate the continuum from the red
side of the Ly$\alpha$ forest, where there is no Ly$\alpha$
absorption.  This strategy has been applied by Press, Rybicki, \&
Schneider (1993), (hereafter PRS93), and by Bernardi et al.  (2003).
The other method is to locate regions in the Ly$\alpha$ forest where
there is apparently no absorption, fit these with a high-order
polynomial, and call this the continuum level (Rauch et al.  1997,
McDonald et al.  2000).

Each approach has systematic difficulties.  The first may cause an
underestimate of the mean transmitted flux: This is because, as SMM03
point out, there may be a break in the slope of the quasar continuum
near the Ly$\alpha$ emission line; the continuum becoming less steep
as one moves across the Ly$\alpha$ line from red to blue, as seen in
low redshift quasar spectra (Zheng et al.  1997, Telfer et al.
2002).\footnote{The exact location of the break is poorly determined,
due to the large number of emission lines near the break point.
Telfer et al.  (2002) indicate only that the break occurs between
$1200 - 1300 \AA$.  On the other hand, Zheng et al.  (1997) found,
with a smaller sample, that the break occurs near $1050 \AA$, which
would be irrelevant for determining the mean flux in the Ly$\alpha$
forest.  The reason for the difference in the position of the break
point between the two studies is not clear.} Furthermore, the slope of
the quasar continuum varies significantly from quasar to quasar
(Telfer et al. 2002). The fitting together of `unabsorbed' regions in
the second method adds an undesirable subjective element to the data
analysis, and inevitably breaks down at sufficiently high redshift
where there are no, or very few, unabsorbed regions in the Ly$\alpha$
forest.  This approach tends to overestimate the mean transmitted flux
in the Ly$\alpha$ forest. There is yet another difficulty with this
type of continuum fitting. The difficulty arises because quasar
spectra are typically taken with an echellograph. The full quasar
spectrum is then pieced together from several individual echelle
orders, with the signal to noise of the spectrum dropping off near the
edges of each echelle order (see e.g. Hui et al. 2001). This imprints
a periodic structure in the noise, an added obstacle in continuum
estimation.

In contrast to the measurements of the mean transmitted flux, the
unabsorbed quasar continuum level does not need to be estimated in
order to measure the flux power spectrum.  In measuring the flux power
spectrum, the power spectrum of the fluctuations in the flux about the
mean, relative to the mean is the quantity of interest.  That is, the
power spectrum of $\delta_f = (f - \langle f \rangle)/\langle f
\rangle$ is estimated.  To determine this quantity observationally,
the mean flux $\langle f \rangle$ (Hui et al.  2001, Croft et al.
2002) can be measured directly, without first rescaling to the
unabsorbed continuum level.  This statistic is sensitive to the {\em
shape} of the continuum, since the continuum, and hence the mean flux,
varies slowly across a quasar spectrum, but it is not sensitive to the
{\em normalization} of the quasar continuum.  It is thereby preferable
to have a method of calibrating the amplitude of the ionizing
background that does not depend on estimating the normalization of the
quasar continuum.

Furthermore, when measurements of the flux power spectrum are combined
with measurements of the mean transmitted flux, the resulting
constraints on the linear matter power spectrum depend sensitively on
the assumed mean transmitted flux (Croft et al.  2002, Zaldarriaga et
al.  2001a, Zaldarriaga et al.  2003, SMM03).  This is because the
effect of changing the amplitude of the ionizing background on the
flux power spectrum is rather degenerate with the effect of varying
the linear matter power spectrum.

Additional information, beyond that contained in the flux power
spectrum, is therefore necessary to simultaneously constrain the
linear matter power spectrum and the amplitude of the ionizing
background.  McDonald et al. (2004b), however, note that variations in
the linear power spectrum and the mean transmission, affect the flux
power spectrum differently at different redshifts. Therefore they can
break the degeneracy that we mention using only measurements of the
flux power spectrum, by virtue of the long redshift span and high
statistical precision of data from the SDSS.  For this to work, they
must assume that the mean transmission and other nuisance parameters
evolve smoothly with redshift. The technique we will describe below
can provide a useful check of these assumptions, and possibly tighten
their constraints further.  For this purpose, an alternative to the
usual source of `additional information', the mean transmitted flux,
is desirable.

Croft et al. (2002) suggest one such alternative method for
calibrating the amplitude of the ionizing background based on
measuring the number of times the flux crosses a given threshold in an
observed quasar spectrum and comparing this with the threshold
crossing frequency in simulated quasar spectra. In this paper, we take
this suggestion a step further by measuring the full probability
distribution function (pdf) of $\delta_f$, which also does not require
an estimate of the unabsorbed continuum level.  {\em We find that this
quantity itself is sensitive to the amplitude of the ionizing
background, and we thereby sidestep one of the biggest loopholes in
parameter estimation from the Ly$\alpha$ forest -- by going straight
from quasar spectra to parameter constraints without continuum
fitting.}

This extends previous work which measured the probability distribution
of flux (normalized to the unabsorbed continuum level) and compared to
simulations. In particular, Rauch et al. (1997) and McDonald et al.
(2000), found that canonical $\Lambda$CDM models provide a good match
to the observed probability distribution of the flux (normalized to
the unabsorbed continuum level), and Gaztanaga \& Croft (1999) studied
the one-point flux pdf in the context of the gravitational instability
model of the Ly$\alpha$ forest.  Desjacques and Nusser (2004) also
recently emphasized the advantage of using the pdf of the flux
(normalized to the unabsorbed continuum level) along with the flux
power spectrum to constrain the linear matter power spectrum.

In the present paper, we first measure the pdf of $\delta_f$, and
demonstrate that this quantity is insensitive to the behavior of the
quasar continuum.  Next, we combine our measurement of the pdf of
$\delta_f$ with the flux power spectrum measurements of Croft et
al. (2002) at $z = 2.72$.  Our intention here is simply to illustrate
that the pdf of $\delta_f$ is useful, when employed in conjunction
with the usual flux power spectrum, for constraining cosmological
parameters and the parameters that describe the physics of the IGM.
Our measurements of the pdf of $\delta_f$ can be combined with more
precise measurements of the flux power spectrum from SDSS data (from
McDonald 2004a, or Hui et al. in prep.).  In addition, the pdf of
$\delta_f$ can itself be well-measured from SDSS data (Burgess, Burles
et al. in prep.).  Measurements of the pdf of $\delta_f$ may be
regarded as part of a larger effort to incorporate higher order
statistics into the analysis of the Ly$\alpha$ forest, as suggested by
Zaldarriaga et al. (2001b), Mandelbaum et al. (2003), and Viel et
al. (2004a). The pdf, as a one point statistic, is a sensible starting
place for incorporating higher-order statistics into the analysis of
the Ly$\alpha$ forest.

The outline of the paper is as follows. In \S \ref{degeneracy} we
illustrate the degeneracy between the amplitude and slope of the mass
power spectrum and the mean transmitted flux in the Ly$\alpha$ forest,
reiterating the main points of SMM03. In \S \ref{pdf_measure} we
present a measurement of the pdf of $\delta_f$ from Keck HIRES data
and show that our measurement is robust to the treatment of the quasar
continuum. In \S \ref{joint_constraints} we illustrate how the
measurement of the pdf of $\delta_f$ helps to tighten constraints on
the slope and amplitude of the mass power spectrum. In \S
\ref{alpha_constrain} we present constraints on the slope of the
temperature-density relation, and in \S \ref{discussion} we conclude.
In the Appendix, we present some details regarding our simulations,
and examine the convergence of our results with respect to simulation
resolution and box size. Tables of the measurements presented in this
paper are available electronically at the website {\em
http://t8web.lanl.gov/people/heitmann/lyma/}.

\section{Degeneracy between the amplitude and slope of the matter
power spectrum and the mean transmitted flux} 
\label{degeneracy}

In this section we illustrate the degeneracy between the amplitude of
the mass power spectrum and the mean flux in the Ly$\alpha$ forest.
We first demonstrate that in the absence of assumptions about the mean
flux in the Ly$\alpha$ forest, only very weak constraints on the
amplitude and slope of the linear mass power spectrum are obtained
(SMM03). We illustrate our points using observational measurements of
the flux power spectrum from Croft et al. (2002) at a redshift of
$\langle z \rangle = 2.72$. In order to introduce notation, we provide
a very brief recap of theoretical models of the Ly$\alpha$ forest.
For more details the reader is referred to e.g, Hui et al. (1997).

Assuming photo-ionization equilibrium, the optical depth to Ly$\alpha$
absorption is given by
\begin{equation}
\tau = A (1 + \delta)^{2 - 0.7 \alpha}.
\label{tau}
\end{equation}
Here $\delta$ is the gas over-density and $\alpha$ is the power law in
the temperature-density relation, i.e., $T = T_0 (1 +
\delta)^\alpha$.\footnote{Here $T$ is the temperature at overdensity
$\delta$, and $T_0$ is the temperature at the cosmic mean density. The
temperature of the low density gas in the IGM at $z \sim 3$ is
expected to be tightly correlated with its over-density (Hui \& Gnedin
1997).} The parameter A is proportional to (e.g., Rauch et al. 1997,
Hui et al. 2001) 
\begin{equation}
A \propto \left(\Omega_b h^2\right)^2 \frac{T_0^{-0.7}}{\Gamma_{\rm HI}}
\frac{H_0}{H(z)} \left(1+z\right)^6,
\label{Adef}
\end{equation}
where $\Omega_b$ is the baryon density in units of the critical
density, $H(z)$ is the Hubble parameter at redshift $z$, $H_0$ is the
Hubble parameter today, $h$ is $H_0/100$ km/s/Mpc, and $\Gamma_{\rm
HI}$ is the photoionization rate of hydrogen, which is directly
proportional to the amplitude of the ionizing background. The
parameter $A$ is determined by a product of cosmological parameters,
which are known to high accuracy, with the combination
$T_0^{-0.7}/\Gamma_{\rm HI}$, which is uncertain. The optical depth is
then shifted into redshift space and convolved with a thermal
broadening window. The flux transmitted through the Ly$\alpha$ forest
is given by $F/F_c = e^{-\tau}$, and the amount of flux absorbed by
the Ly$\alpha$ forest is $1 - F/F_c$. Here $F/F_c$ denotes the flux
normalized to the unabsorbed quasar continuum level. Our model of the
Ly$\alpha$ forest is then complete given the baryonic matter density
and peculiar velocity fields specified by our numerical simulations
for a given cosmological model (see the Appendix for details regarding
simulations).\footnote{We generate the baryonic density field using an
implementation of HPM, the pseudo-hydrodynamic recipe proposed by
Gnedin \& Hui (1998).}

Our methodology for obtaining constraints on the matter power spectrum
and the physics of the IGM from this modeling is very similar to that
of Zaldarriaga et al. (2001a, 2003) (see also Lidz et al. 2003). In
brief, we generate the flux power spectrum and pdf for a large grid of
simulated models, calculate the likelihood of the model given the
data, and marginalize over nuisance parameters to obtain constraints
on the amplitude and slope of the linear matter power spectrum.  For
this purpose, we ran a grid of models describing the IGM, with
parameter vector $(a, n, T_0, \alpha, \langle F/F_c \rangle)$. Our
grid covers the following range in each parameter:

\begin{itemize}

\item $a=(0.0919,0.1301,0.1479,0.1682,0.1914,0.2178,\\ 
0.2480,0.2827,0.3227,0.3691,0.4236,0.5327)$

\item $n=(0.7,0.8,0.9,1.0,1.1,1.2,1.3)$

\item $T_0=(200,250,300,350,400, \\
450)$ (km/s)$^2$

\item $\alpha=(0.0,0.1,0.2,0.3,0.4,0.5,0.6)$

\item $\langle F/F_c
\rangle=(0.670,0.680,0.685,0.690,0.695,0.700,\\
0.705,0.710,0.715,0.720,0.725,0.730,0.735,0.740,\\
0.745,0.750,0.755,0.760,0.770,0.780)$

\end{itemize}

The parameters of our simulated model are : $a$, the scale factor of
the simulation output, which effectively corresponds to the
normalization of the matter power spectrum; $n$, the slope of the
primordial power spectrum, each slope corresponding to a different
linear power spectrum slope on the scales probed by the Ly$\alpha$
forest; $T_0$, the temperature at mean density; \footnote{The
temperature is given in units of (km/s)$^2$, which is related to the
temperature in K by $T_0 = T_0[(km/s)^2]\times 10^4/165$ K.}
$\alpha$, the slope of the temperature density relation and $\langle
F/F_c \rangle$, the mean transmission in the Ly$\alpha$ forest. In
this paper we generally parameterize our model in terms of the mean
transmission, $\langle F/F_c \rangle$, although varying this parameter
corresponds to varying $A$ and hence the amplitude of the ionizing
background [Eqn.~(\ref{Adef})]. We emphasize that, in later parts of
the paper, we do not use measurements of the mean transmission to fix
the amplitude of the ionizing background. In spite of this we will
still parameterize our model in terms of $\langle F/F_c \rangle$,
rather than in terms of the amplitude of the ionizing background. This
will be convenient for comparing our results with different
measurements of $\langle F/F_c \rangle$ that appear in the
literature. In addition to these parameters, we need to specify a
thermal history in order to calculate the gas pressure term in our HPM
simulations. We describe this in the Appendix.

In this paper we use large volume simulations, which lack sufficient
resolution to completely resolve the small-scale structure in the
Ly$\alpha$ forest.  Hence, in comparing our simulated models of the
flux power spectrum with Croft et al. (2002)'s measurements, we
confine the comparison to scales of $k \lesssim 0.02$
s/km.\footnote{Our numerical simulations were run primarily for
comparison with quasar spectra from the SDSS. These spectra have
relatively poor spectral resolution, motivating us to sacrifice
simulation resolution for simulation volume.} In the Appendix, we
demonstrate that even restricting our analysis to these scales our
simulation resolution is somewhat marginal, and we attempt to quantify
the resulting systematic error. We will generally quote constraints on
the amplitude and slope of the matter power spectrum as constraints on
the dimensionless amplitude of the power spectrum at redshift $z$ and
scale $k_p=0.03$ s/km, $\Delta^2=k_p^3 P(k_p,z)/2 \pi^2$, and the
logarithmic slope at the same scale and redshift, $n_{\rm eff} = d
{\rm ln} P(k_p)/d {\rm ln} k$ (Croft et al. 2002). For reference, at
$z = 2.72$, the central redshift of the data sample in Croft et
al. (2002), the scale $k_p$ corresponds to a co-moving scale of $k_0 =
H(z = 2.72)/(1+z) k_p$, which is $k_0 = 3.24 h$ Mpc$^{-1}$ for a flat,
$\Lambda$CDM cosmology with matter density $\Omega_m=0.3$.

Next, we have to consider the priors we should adopt on the different
nuisance parameters involved in the modeling. For the most part, we
will be fairly conservative in adopting these priors. The temperature
of the IGM, however, has been constrained using simulations with
smaller volume, but higher resolution than our present simulations
(see, e.g., McDonald et al. 2001, Zaldarriaga et al. 2001a). The most
pronounced effect of gas temperature on the flux power spectrum is a
small-scale smoothing on scales of $k \gtrsim 0.05$ s/km. Since these
scales are not fully resolved by our simulations, we will not attempt
to constrain $T_0$ with the present simulations.  Motivated by results
from the higher resolution simulations, however, we adopt a prior on
the temperature of the IGM\footnote{In the present paper, we limit our
theoretical modeling to modeling at a redshift of $z = 2.72$. This
prior, and those listed subsequently, thus refer to the priors that we
adopt at $z=2.72$.} of $T_0=300 \pm 50$ (km/s)$^2$. (We will relax our
prior on $T_0$ in \S \ref{alpha_constrain} to demonstrate that our
constraint on the temperature-density relation, $\alpha$, is not
sensitive to the prior on $T_0$.)  This prior on the temperature
corresponds to the constraint found by McDonald et al. (2001), except
with an error bar approximately twice as large.  Conservatively, we
will let $\alpha$ vary freely over the entire physically allowed range
of $\alpha = 0.0 - 0.6$ (Hui \& Gnedin 1997).  There is one sense in
which we are not conservative, however: in calculating the gas
pressure term in our HPM simulations, we fix the thermal history of
the IGM (see the Appendix for details). This was done partly to reduce
computational burden, but also because, on the large scales we
consider here ($k \lesssim 0.02$ s/km), the effects of gas pressure
smoothing are likely to be sub-dominant compared to the size of Croft
et al. (2002)'s error bars.

\begin{figure}[t]
\vbox{ \centerline{
\epsfig{file=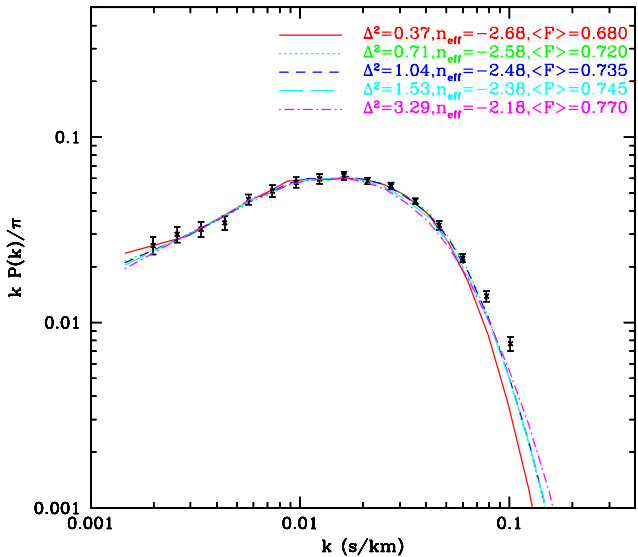,width=9.5truecm}} 
\figcaption[]{Auto spectrum from Croft et al. (2002) at
$\langle z \rangle = 2.72$ compared to models that  compensate for
their very different mass power spectrum amplitudes and slopes by
having different mean transmissions. In our subsequent analysis, we
only compare theoretical models with data on scales of $k$ less than
$0.02$ s/km owing to our limited simulation resolution. The models are
completely specified by the  parameter vector $(\Delta^2, n_{\rm eff},
T_0, \alpha, \langle F/F_c \rangle)$. The models shown are specified
by the following parameter sets: $(0.37, -2.68, 350, 0.5, 0.680)$,
$(0.71, -2.58, 300, 0.4, 0.720)$, $(1.04, -2.48, 300, 0.4, 0.735)$,
$(1.53, -2.38, 300, 0.4, 0.750)$, $(3.29, -2.18, 300, 0.3, 0.770)$. 
\label{deg}} }
\end{figure}

With this in mind, we generate the flux auto spectrum for each model
in our grid and compute the likelihood of the model given the data. We
show the resulting likelihood contours in \S \ref{joint_constraints}.
Presently, we illustrate the degeneracy between $\langle F/F_c
\rangle$ and the amplitude and slope of the mass power spectrum, with
example model fits from our grid. The model fits each have very
different mass power spectrum amplitudes and slopes, compensated by
rather different values of the mean flux, and somewhat by small
changes in the other nuisance parameters. This is shown in
Figure~\ref{deg}.  In this plot, one can see that, assuming nothing
about the mean flux in the Ly$\alpha$ forest, the amplitude of the
mass power spectrum is uncertain by an order of magnitude, and the
slope of the mass power spectrum has an error of at least
$20\%$. While the range of values for the mean transmitted flux
considered in the figure is perhaps excessively conservative, a wide
range of mean transmitted flux priors have been considered in the
literature. Specifically, the prior on the mean transmission adopted
by Croft et al. (2002), is $\langle F/F_c \rangle = 0.705 \pm 0.012$,
while SMM03 consider two priors based on the measurements of McDonald
et al. (2000): $\langle F/F_c \rangle = 0.742 \pm 0.012$, and one with
expanded error bars, $\langle F/F_c \rangle = 0.742 \pm 0.027$, and
Viel et al. (2004b) adopt $\langle F/F_c \rangle = 0.730 \pm
0.011$.\footnote{The Croft et al. (2002) prior is based on
measurements by PRS93, who give a power law fit to the effective
optical depth, $\tau_{\rm eff} = -ln \langle e^{-\tau} \rangle = A
(1+z)^{\gamma +1}$, with error bars of $A = 0.0175 - 0.0056 \gamma \pm
0.0002$ and $\gamma = 2.46 \pm 0.37$. It is not exactly clear how to
interpret these error bars: there has been some debate over what the
PRS93 fit implies for the central value of $\langle F/F_c \rangle$ and
its error bar. The prior adopted by Croft et al. (2002) corresponds to
using the central values of $A$ and $\gamma$ to determine the central
value of $\langle F/F_c \rangle$, but the rationale behind their
assumed error bar on $\langle F/F_c \rangle$ is unclear.  SMM03, on
the other hand, argue that assuming that $A$ and $\gamma$ are
Gaussian-distributed actually implies a larger central value for
$\langle F/F_c \rangle$.  Meiksin \& White (2004), with a still
different interpretation of the same error bars, also argue for a
larger central value for $\langle F/F_c \rangle$. SMM03 suggest that
their interpretation of the PRS93 measurements brings them in closer
agreement with measurements from McDonald et al. (2000) (see above for
the mean transmitted flux that SMM03 adopt).  However, in addition to
the uncertainty involved with interpreting the PRS93 measurements,
there are further reasons to be hesitant about concluding that the
different measurements of $\langle F/F_c \rangle$ are consistent with
each other.  First, as we mentioned earlier, the same authors argue
that the PRS93 measurements, by extrapolating the unabsorbed continuum
level from the red side of the Ly$\alpha$ emission line, should be
biased low due to a break in the slope of the continuum near the
Ly$\alpha$ emission line.  Second, Bernardi et al. (2003) use a method
similar to that of PRS93, (yet more sophisticated), to estimate
$\langle F/F_c \rangle$ and find $\langle F/F_c\rangle \sim 0.70$ at
$z \sim = 2.72$, with $\sim 2 \%$ error bars. Presently we remain
agnostic as to which measurement of $\langle F/F_c \rangle$ is most
accurate.  Our goal is just to illustrate the sensitivity of matter
power spectrum constraints to different assumptions about the mean
flux that have appeared in the literature.  In \S
\ref{joint_constraints}, however, we obtain constraints on the mean
flux in the Ly$\alpha$ forest using our measurements of the pdf of
$\delta_f$.}  The best fit amplitude of the mass power spectrum with
the Croft et al. (2002) prior on the mean transmitted flux differs by
a factor of $\gtrsim 2$ than that obtained with the SMM03 prior (see
the models in Figure~\ref{deg}).  Furthermore, SMM03 show that the
error contours become very large with their prior on the mean
transmitted flux, in comparison with the error contours corresponding
to the Croft et al. (2002) prior. The flux field becomes primarily
sensitive to larger and larger over-densities as the mean transmitted
flux increases, and one loses sensitivity to the {\em linear} matter
power spectrum (see also Zaldarriaga et al.  2003).  The figure also
illustrates that including information at high $k$, (we only compare
theory with data on scales of $k \lesssim 0.02$ s/km due to the
limited resolution of our simulations), would only help slightly in
obtaining tighter constraints.

The lesson here, reiterating the main point of SMM03, as well as
cautions mentioned by Croft et al. (2002), is that one needs a tight
constraint on the mean transmission in the Ly$\alpha$ forest in order
to derive constraints on the mass power spectrum from the Ly$\alpha$
forest flux power spectrum. There are a few possible ways to
circumvent this difficulty.  One approach is to determine any bias in
estimates of the mean absorption obtained by fitting together
`unabsorbed' regions in the Ly$\alpha$ forest using simulated spectra,
as Rauch et al. (1997) and McDonald et al. (2000) attempted to do. An
inherent difficulty is that mock spectra are typically produced in
simulation boxes that are small compared to the length of spectra over
which the continuum-fitting procedure is performed.  Furthermore, the
size of the estimated bias may be model-dependent. Another approach
might be to correct estimates of the unabsorbed quasar continuum level
extracted from the red side of Ly$\alpha$ based on the break found in
low redshift quasar spectra. However, the correction depends on the
position of the break, which is poorly determined from the low
redshift quasar spectra.  As we mentioned in the Introduction,
McDonald et al. (2004b) circumvent the degeneracy we mention by
exploiting the fact that the flux power spectrum depends differently
on the modeling parameters at different redshifts. The disadvantage of
this approach is that they must assume that the nuisance parameters
evolve smoothly with redshift.  Yet a different approach, which we
pursue presently, is to avoid using the mean transmitted flux at all,
and to instead calibrate the amplitude of the ionizing background, [or
equivalently the parameter $A$ in Eqn.~(\ref{Adef})], using a
different statistic entirely.

\section{Measuring the probability distribution of the flux field 
and its error bars}
\label{pdf_measure}

In this section, we present a measurement of the probability
distribution of the flux field, $\delta_f$, which will be useful in
calibrating the parameter $A$ in Eqn.~(\ref{Adef}), as we will
demonstrate. We choose a particular estimator for $\delta_f$ that we
can measure robustly from the data sample as well as the numerical
simulations. Our objectives here are to: 1) Choose a quantity that can
be measured accurately from our simulations, 2) Use an estimator of
$\delta_f$ that does not rely on estimating the level of the
unabsorbed quasar continuum, 3) Identify an estimator that is
insensitive to contamination from the unknown {\it shape} of the
quasar continuum.

Several different estimators for $\delta_f$ have been discussed in the
literature in the context of measuring the flux power spectrum (Hui et
al.  2001, Croft et al.  2002).  We now briefly describe these
estimators.  The number of photon counts, $N_p$, in the Ly$\alpha$
forest portion of a quasar spectrum is the product of a slowly varying
quasar continuum, $N_c$, with $e^{-\tau}$, (ignoring additional
opacity from metal absorption lines).  Fluctuations in $N_p$ arising
from fluctuations in $e^{-\tau}$ have to be separated out from those
that arise due to structure in the quasar continuum.  This can be
accomplished, at least on sufficiently small scales, since the quasar
continuum, $N_c$, varies smoothly in comparison to $e^{-\tau}$.  We
proceed to discuss how this is done.  The mean transmission, i.e.,
$e^{-\tau}$ averaged over an ensemble of quasar spectra, $\langle
F/F_c \rangle = \langle e^{-\tau} \rangle$, is not a function of
wavelength for a sufficiently small stretch of a quasar spectrum.  In
spite of this, the mean number of photon counts depends on wavelength
due to the slowly varying structure in the quasar continuum.  This
`running mean' quasar count can be estimated from the spectrum by
fitting a low order polynomial to the quasar spectrum (Hui et al.
2001), or by smoothing the spectrum with a large radius filter (Croft
et al.  2002).  Then $\delta_f$ is formed by subtracting the running
mean quasar count from the raw quasar count, finally dividing out by
the running mean.  This procedure thereby offers a means of forming
$\delta_f$ and measuring its flux power spectrum without first
estimating the normalization of the quasar continuum.

The estimator we use for $\delta_f$ in measuring the flux pdf is very
similar to the estimator used by Croft et al. (2002) for measuring the
flux power spectrum.  In particular, we will demonstrate that
$\delta_f$ defined by:
\begin{equation}
\delta_f(\lambda) = \frac{f_{r}(\lambda) -
f_{R}(\lambda)}{f_{R}(\lambda)} 
\label{deltaf} 
\end{equation}
satisfies the three desired criteria we outline above.  In this
equation $f_{r}(\lambda)$ indicates the value of the flux field, $f$,
at wavelength $\lambda$ when smoothed with a Gaussian filter of radius
$r$ -- i.e., the variance of the Gaussian is $\sigma^2 = r^2$, and
$f_{R}(\lambda)$ denotes the flux field at wavelength $\lambda$ when
smoothed with a Gaussian filter of radius $R$.  The filter, $r$, has a
small radius, and serves to take out small-scale power on scales that
are under-resolved in our simulation. In the Appendix we demonstrate
that $r = 30$ km/s is an appropriate filter for our present
simulations. The filter, $R$, represents a large-scale smoothing that
defines the running-mean flux in the quasar spectrum.  Given a sample
of quasar spectra it is straightforward to estimate $\delta_f$ as
defined above, and measure its one-point probability distribution
function.

We carry out this procedure using spectra obtained with the High
Resolution Echelle Spectograph~(HIRES) on the Keck telescope. These
quasar spectra are the same as described in Rauch et al. (1997) and in
McDonald et al. (2000). (See these papers for references regarding the
data reduction procedure.) The resolution of the quasar spectra is
FWHM $\sim 6.6$ km/s and the typical signal to noise per $0.04 \AA$
pixel is $S/N \sim 50$. The quasar spectra, listed by name, emission
redshift, and wavelength range examined, are shown in
Table~\ref{lambda_use}.  Before measuring the flux pdf, but after
performing the smoothing procedure described by Eqn.~(\ref{deltaf}),
we attempt to cut out metal lines, damped-lyman alpha systems, and
spurious pixels, using wavelength cuts similar to those in Rauch et
al. (1997) and McDonald et al.  (2000). Unlike Rauch et al. (1997) and
McDonald et al. (2000), we use only the `raw' data, as opposed to
continuum-fitted data, in making our measurements.

We split the data sample into four relatively narrow redshift bins,
and measure the flux pdf in each redshift bin. The redshift bins are
defined in Table~\ref{zbins} below. One of the redshift bins is chosen
to have a mean redshift of $z=2.72$, the central redshift of the Croft
et al. (2002) sample. In the present paper, we will limit our
theoretical analysis to this redshift bin, and we will use it to
illustrate our procedure for measuring the flux pdf.
\begin{deluxetable}{ccc}
\tablecaption{Redshift Bins for PDF measurements
\label{zbins}}
\tablehead{
\colhead{$z_{\rm low}$} &
\colhead{$z_{\rm high}$} &
\colhead{$\langle z \rangle$}
}
\startdata
2.20    &       2.45    &       2.26\\

2.46    &       3.00    &       2.72\\

3.01    &       3.70    &       3.28\\

3.71    &       4.43    &       3.99\\
\enddata
\end{deluxetable}

After splitting our data sample into several redshift bins, we smooth
each quasar spectrum in order to form $\delta_f$ as described above,
and then re-bin each spectrum into pixels of size $20$ km/s. Data
within $2,000$ km/s, (which corresponds to $4$ times the radius of our
large-scale smoothing filter, $R$), of the edge of each spectrum are
cut after performing the smoothing to avoid edge effects.  Upon
forming the flux field from each quasar spectrum, we measure the pdf
using $36$ bins in $\delta_f$, each with a width of $0.061538$. The
center of the lowest flux bin corresponds to $\delta_f = -1.01538$. In
the event that a pixel has flux lower (higher) than the lower (upper)
edge of the smallest (largest) $\delta_f$ bin it is included in the
lowest (highest) $\delta_f$ bin.  We compute the average flux in each
bin, and use this, rather than the bin center, when comparing with
theoretical models.  
as 
allows 

Next, we discuss our procedure for estimating error bars.  In order to
measure the variance and covariance of our estimates of the pdf of
$\delta_f$, we use a jackknife technique.  Specifically, we 1)
estimate the pdf from the full data sample, 2) divide the data set
into $n_g = 30$ different subgroups, and 3) we estimate the pdf of the
data sample {\em omitting} each sub-group. Let $\hat{P}(\delta_i)$
represent the pdf estimated from the full data sample for a bin with
average flux, $\delta_i$. In addition, let $\tilde{P}_k(\delta_i)$
represent the same quantity, estimated not from the full data sample,
but from a data sample that omits the pixels in the kth subgroup. Then
the jackknife estimate of the covariance between the estimates of the
pdf in a bin with average flux, $\delta_i$, and a bin with average
flux, $\delta_j$ is given by:
\begin{equation}
Cov(i,j) = \sum_{k=1}^{n_g} \left[\hat{P}(\delta_i) -
\tilde{P}_k(\delta_i)\right]\left[\hat{P}(\delta_j) -
\tilde{P}_k(\delta_j)\right] 
\label{coveq}
\end{equation}
and the diagonal variance, for a bin with flux $\delta_i$, is
$\sigma^2_i = Cov(i,i)$.  

\begin{figure}
\plotone{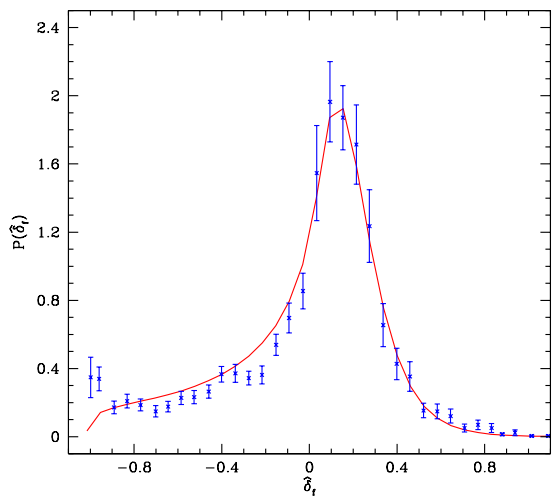}
\caption{A demonstration that our lognormal model roughly reproduces
the pdf of the data.  \label{logn_model}}
\end{figure}

In order to check the accuracy of our jackknife estimate of the error
bars, we generate mock spectra using the lognormal model described in
McDonald et al. (2004a).\footnote{The lognormal model we adopt is
slightly different than that of McDonald et al. (2004a). Specifically,
McDonald et al. (2004a) go from the lognormal `density field',
$\Delta$, to an optical depth field with the transformation $\tau
\propto \Delta^2$. Here we instead adopt $\tau \propto\Delta^{1.79}$,
and a slightly different proportionality constant to relate optical
depth to density. If the temperature-density relation is $T=T_0 (1 +
\delta)^\alpha$, our choice corresponds to $\alpha=0.3$, while
McDonald's choice corresponds to the isothermal case, $\alpha=0.0$. We
find that this provides a better fit to the pdf, while retaining a
rough match to the observed redshift evolution of the mean transmitted
flux in the Ly$\alpha$ forest.}  This provides a test of our error
estimation procedure: first we measure the dispersion across many
independent realizations generated using our lognormal simulations,
and second we make a jackknife estimate from a single realization. To
the extent that the jackknife method agrees with error bars estimated
from many independent realizations, we can be confident that it is
sound.  We illustrate our procedure for constructing mock realizations
of the data considering the redshift bin centered on $\langle z
\rangle = 2.72$ as an example.  The corresponding procedure for the
other redshift bins is very similar.  First we generate four lines of
sight, corresponding to the number of lines of sight in our actual
data sample that fall into the $\langle z \rangle = 2.72$ bin. Each
line of sight spans the length of our redshift bin, $z = 2.46 - 3.00$,
on a fine grid of $16384$ pixels.  We form $\delta_f$ according to
Eqn.~(\ref{deltaf}), coarse-sample onto pixels of size $20$ km/s, and
cut out regions of spectra until the mock spectra have exactly the
same wavelength coverage as the actual spectra. In
Figure~\ref{logn_model}, we show a measurement of the pdf from many
realizations of our lognormal model as compared to a measurement of
the pdf from real data. The comparison illustrates that the lognormal
model roughly reproduces the pdf of the actual data, although there
are some differences between the model and data.  We emphasize that we
only use the lognormal model to 1) test our jackknife estimates of the
variance of the pdf, 2) indicate the correlation coefficient between
our estimates of the pdf in different flux bins. This second step is
necessary because, as we detail below, our estimates of the
off-diagonal elements of the covariance matrix from our present data
set are too noisy to be useful. In the future, with larger data
samples, we can just estimate the covariance matrix directly.

\begin{figure}
\plotone{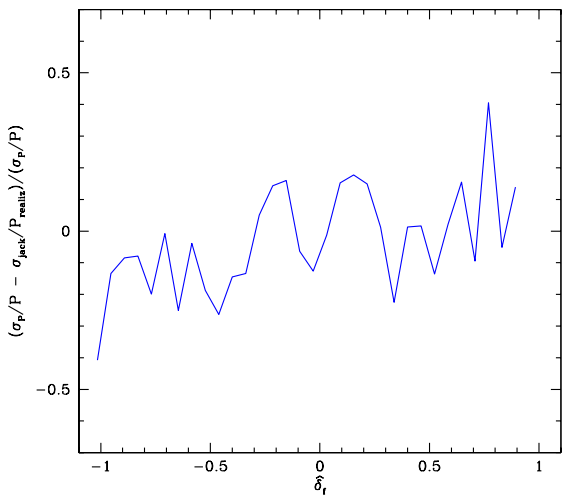}
\caption{\label{errtest} Comparison of two estimates of the error bars
on the flux pdf, as measured from lognormal simulations.  The first
error bar estimate comes from the dispersion across many independent
realizations, and the second is a jackknife estimate.  The fractional
difference between the fractional error bars is shown.  }
\end{figure}
How well does the jackknife technique described above work?  In
Figure~\ref{errtest} we compare our jackknife estimate of the
fractional error on the flux pdf with an estimate from the dispersion
across many simulation realizations.  The jackknife error estimate is
generally good to better than $20\%$, but seems to produce a
systematic overestimate of the error at low $\delta_f$. The apparent
overestimate of the errors is at least conservative in the sense that
parameter constraints would only tighten if we were to adopt smaller
error bars on these points.  On the other hand, the error bars on the
points in the high $\delta_f$ tail may be {\em underestimated} using
the jackknife estimator.  In this case, there are so few pixels in the
high $\delta_f$ tail of the pdf that our jackknife method is
unreliable. To overcome this difficulty, we again adopt a conservative
approach and use the variance measured from the lognormal simulations
to represent the variance for the high $\delta_f$ bins in our
measurement from real data.

\begin{figure} \plotone{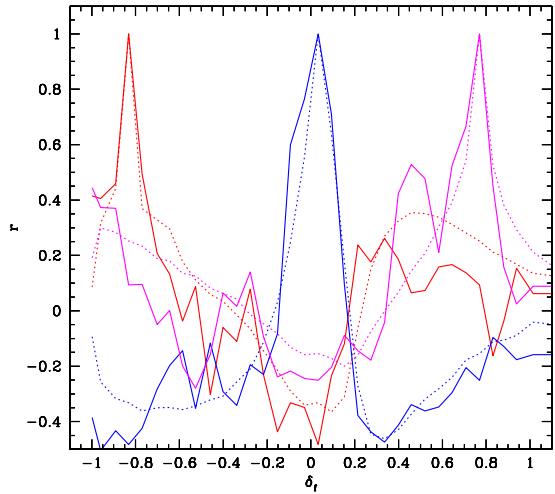}
\caption{\label{corrco_comp}
We show, on the one hand the noisy jackknife estimates of the
correlation coefficient from the data (solid lines), and on the other
hand, simulation estimates of the correlation coefficient (dotted
lines). The red line is the correlation coefficient for a bin with an
average flux of $\delta_f=-0.832$, the blue line has an average flux
of $\delta_f=0.0326$, and the magenta line $\delta_f=0.769$. The
simulation measurements are fairly consistent with a smooth version of
the noisy jackknife estimate from real data. }
\end{figure}

Next, we consider the off-diagonal terms in the covariance matrix. We
find that our jackknife estimates of the off-diagonal elements of the
covariance matrix are quite noisy. After all, we are attempting to
measure $36 \times 37 / 2 = 666$ matrix elements from only $3499$ data
pixels. To get around this difficulty, we first measure the
correlation coefficient from many realizations of the mock lognormal
spectra. The correlation coefficient is defined by $r(i, j) =
Cov(i,j)/\sqrt{Cov(i,i) Cov(j,j)}$.  Next we assume that {\em the real
data has the same correlation coefficient as the mock data}, and
estimate the data covariance from the simulated correlation
coefficient $r(i,j)$ and the jackknife estimate of the data variance,
i.e.  $Cov_{\rm d}(i,j) = r_{\rm s}(i,j) \sqrt{Cov_{\rm d}(i,i)
Cov_{\rm d}(j,j)}$.  Here the subscript `d' indicates a quantity
measured from real data, while the subscript `s' represents a quantity
measured from the lognormal simulations. This approach allows us to
estimate the off-diagonal terms in the data covariance matrix, which
are significant, but cannot be estimated cleanly from the data
alone. The approach is justified to the extent that the correlation
coefficient in the lognormal simulation is representative of the true
correlation coefficient in the real data.  In Figure~\ref{corrco_comp}
we show a comparison of the correlation coefficient measured from the
data with a jackknife estimator, which is noisy, and that measured
from the lognormal simulations, which is smooth. From the figure, one
can see that the simulation measurement is fairly consistent with a
smooth version of the noisy jackknife estimate, justifying our
approach.

This procedure then provides us with an estimate of the full
covariance matrix. There is one last detail that we need to address
before proceeding with the remaining analysis. The difficulty is that
the flux pdf is required to satisfy a normalization condition, $\int
d\delta_f P(\delta_f) = 1$, and the data vectors are hence {\em not
linearly independent}, and the covariance matrix is nearly singular.
We therefore diagonalize the covariance matrix, and rebuild it by
eliminating the eigenvector with the smallest eigenvalue. In other
words, the rebuilt covariance matrix can be written as
\begin{equation}
\label{covtrunc}
\tilde{C}_{ij} = \sum_{k=1}^{N_m} U_{ik} U_{jk} \lambda_k,
\end{equation}
where $\tilde{C}_{ij}$ is the `regularized' covariance matrix, $U$ is
the unitary matrix that diagonalizes the covariance matrix, and
$\lambda_k$ is the $k$th eigenvalue of the covariance matrix.  The
eigenvalues are sorted by size, with $\lambda_1$ denoting the largest
eigenvalue. The sum extends up to $N_m=35$, i.e. -- $N_m$ is the
number of bins minus $1$. This `regularized' covariance matrix can
then be stably inverted and used to compute $\chi^2$ when comparing
theoretical models with data (see \S \ref{joint_constraints}).

\begin{figure}[t]
  \vbox{ \centerline{ \epsfig{file=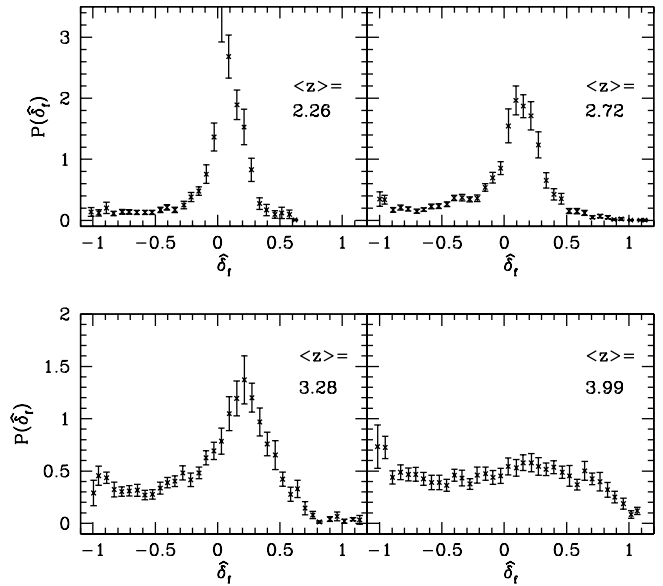,width=10.0truecm}}
    \figcaption[]{Measured flux pdf in redshift bins with mean
redshifts of $\langle z \rangle = 2.26, 2.72, 3.28$, and $3.99$.  
\label{pdf_measured}}}
\end{figure}

Our estimates of the flux pdf and its error bars in the fiducial
redshift bin are given in Table~\ref{pdftable}.  The flux pdf and
error bars for the other redshift bins (see Table~\ref{zbins}) are
given in Tables~\ref{pdf_bin1}--\ref{pdf_bin4}.  In
Figure~\ref{pdf_measured} we show our pdf measurement in the different
redshift bins.  The pdf evolves as one expects -- a larger fraction of
pixels in the spectrum become opaque with increasing redshift. Tables
with the covariance matrices for each redshift bin can be obtained
electronically from the website {\em
http://t8web.lanl.gov/people/heitmann/lyma/}.

With our estimates of the flux pdf and its covariance matrix in hand,
we now demonstrate that our estimator satisfies the criteria we
described in the beginning of this section. The first criterion, that
the estimator be insensitive to the numerical resolution limits of the
simulations, is established in the Appendix. The second criterion is
clearly satisfied, since the estimator defined in Eqn.~(\ref{deltaf})
does not refer to the unabsorbed continuum level.

The third criterion, that our estimator is insensitive to the {\it
  shape} of the quasar continuum, is not obvious.
In order to investigate this, we estimate the pdf of $\delta_f$ from
the `raw' data as well as from a version of the data that has first
been continuum-normalized, as described in Rauch et al. (1997).  If
our procedure of applying a large-scale filtering to form $\delta_f$
is robust to the unknown continuum shape, then we should get very much
the same pdf whether we measure the pdf using the raw data or the
continuum-fitted data.

\begin{figure}[t] \vbox{ \centerline{
\epsfig{file=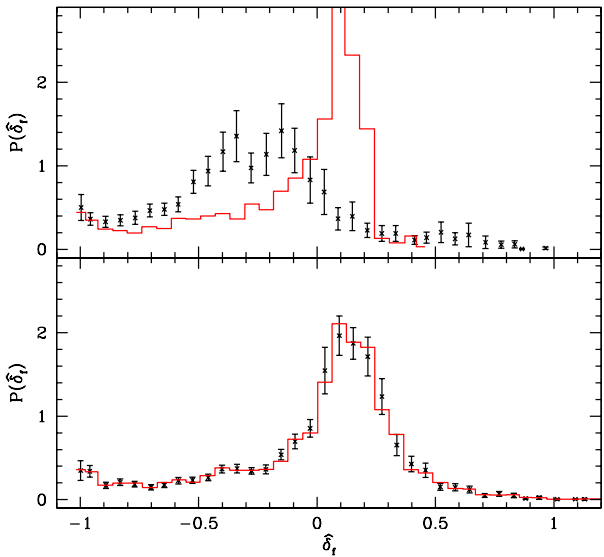,width=9.5truecm}}
\figcaption[]{Flux pdf from continuum fitted data (red
histogram) and from raw data (black points with error bars). The flux
pdfs in the top panel are formed assuming a flat mean to estimate
$\delta_f$, while in the bottom panel $\delta_f$ is defined using a
large-scale smoothing of $R=500$ km/s. The error bars on the pdf
measured from the continuum fitted data, which are comparable to those
from the raw data, are not shown for visual clarity.  The close
agreement between the pdf of the flux field measured from the raw data
and that measured from the continuum-fitted data, when both are
smoothed on a scale of $R=500$ km/s, illustrates that our procedure is
robust to the continuum. The disagreement between the flux pdfs in the
top panel, formed assuming a flat mean, is due to large-scale power in
the continuum.  \label{pdf_robust}}} 
\end{figure}

We illustrate our approach in Figure~\ref{pdf_robust} where we measure
the pdf of $\delta_f$, for each of the raw data and the
continuum-fitted data, in two different ways.  First, we measure the
pdf from each of the raw and the continuum-fitted data, assuming a
flat mean. In this case the pdf of $\delta_f$ differs substantially
between the continuum-fitted and the raw data. The continuum has power
on large scales, and the flux pdf will differ significantly between
the raw and the continuum-normalized data if we don't account for
this. Second, we measure the pdf from each of the raw and the
continuum-fitted data, with $\delta_f$ formed using the $R=500$ km/s
smoothing to define $\delta_f$ as described above. In this case the
pdf of $\delta_f$ is extremely similar between the
continuum-normalized and raw data. In one case we have measured the
pdf of $\delta_f$ from data that has first been continuum-normalized,
and in the second case we did not perform any continuum-normalization,
effectively assuming that the continuum is flat. The similarity
between the pdf of $\delta_f$ under these two very different
assumptions regarding the true quasar continuum, demonstrates that
{\em the pdf of $\delta_f$ is insensitive to our assumptions about the
quasar continuum}.  To provide a quantitative measure of the
similarity between the pdf measured in the two different ways, we
compute $\chi^2$ between the two probability distributions. We find,
comparing our $36~\delta_f$ bins, and including estimates of the
off-diagonal elements in the covariance matrix, that $\chi^2=8.3$.
The pdf of $\delta_f$ for the continuum-fitted and raw data are
therefore in close agreement.\footnote{One might worry that this value
of $\chi^2$ is actually too small. However, one should keep in mind
that we are comparing two probability distributions measured from the
same data, yet with different data reduction procedures.}

\section{Constraints on the Mass Power Spectrum using the flux power
spectrum and the flux pdf}
\label{joint_constraints}

The next step, after demonstrating that we can measure the flux pdf
robustly, is to show that the flux pdf is useful in constraining the
mass power spectrum, when employed in conjunction with the flux power
spectrum. In particular, we emphasize that the flux pdf is sensitive
to $A$, the parameter related to the amplitude of the ionizing
background, and hence we can place constraints on the mass power
spectrum, without making any assumptions about the mean flux in the
Ly$\alpha$ forest.

\begin{figure}[t] \vbox{ \centerline{
        \epsfig{file=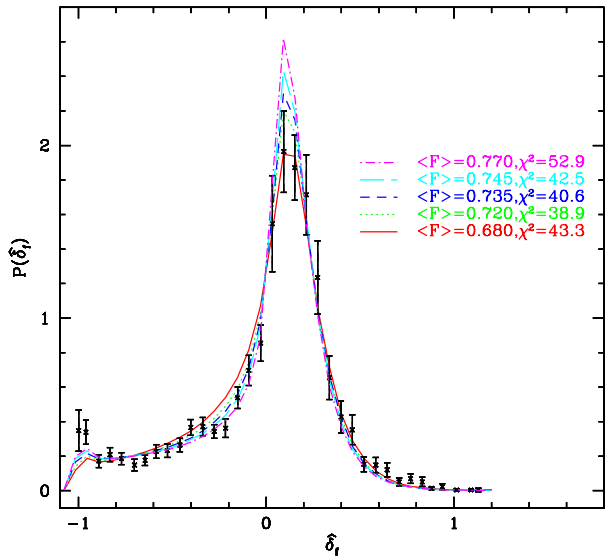,width=9.5truecm}}
\figcaption[]{Flux pdf of the models shown in
Figure~\ref{deg}, illustrating that the flux pdf is helpful in
breaking the degeneracy between the mean transmission and the
amplitude/slope of the mass power spectrum. One should keep in mind
that the error bars in different $\delta_f$ bins are
correlated. \label{pdf_dbreak}} } 
\end{figure}

In Figure~\ref{pdf_dbreak} we show our pdf measurement at $z=2.72$ as
well as theoretical predictions for the pdf for the models shown in
Figure~\ref{deg}, each of which fits the observed flux power spectrum,
in spite of having very different matter power spectra.  While the
flux pdfs of each model look fairly similar by eye, one can
distinguish between them statistically.  Of these models, the minimum
$\chi^2$ obtained from fitting to the PDF alone, occurs at $\langle
F/F_c \rangle = 0.720$, with $\chi^2=38.88$. The fit is reasonable
given that we compare the theoretical flux pdf with the measured flux
pdf for $\sim 34$ degrees of freedom (36 data points - 1 normalization
condition - 1 prior on $T_0$), and so the fit roughly corresponds to
$\chi^2/\nu \sim 1.1$, which should be exceeded randomly $\sim 26 \%$
of the time. The model with a lower mean flux of $\langle F/F_c
\rangle = 0.680$ is a significantly worse fit with
$\chi^2=43.26$. Similarly models with significantly higher mean
transmitted flux are also poor fits to the data. Specifically, the
model with $\langle F/F_c \rangle = 0.750$ has $\chi^2=42.46$, and the
model with $\langle F/F_c \rangle = 0.770$ has $\chi^2=52.86$.

The key point is that it is easy to find two models that, while
differing significantly in their photo-ionizing backgrounds, yield the
same flux power spectrum (and hence have the same variance) -- however
these two models will generally differ in their higher moments.  In
particular, the model with $\langle F/F_c \rangle = 0.680$ has a
variance of $\langle \delta_f^2 \rangle = 0.134$ and a skewness of $S
= \langle \delta_f^3 \rangle/ \langle \delta_f^2 \rangle ^{3/2} =
-0.975$.  The model with $\langle F/F_c \rangle = 0.770$ has a nearly
identical variance, but its skewness is rather different, $S = -1.24$.
To clarify the situation, let us consider a simplified argument for
how the information in the skewness complements the information in the
flux power spectrum.  The pdf of $\delta_f$ is mainly described by two
parameters, its variance $\langle \delta_f^2 \rangle$, and its
skewness, $S = \langle \delta_f^3 \rangle/ \langle \delta_f^2 \rangle
^{3/2}$.  The underlying model, on the other hand, mainly depends on
the variance of the density field, smoothed on some scale, $\langle
\delta_\rho^2 \rangle$, and the amplitude of the ionizing background,
parameterized by $A$, or effectively $\langle F/F_c \rangle$.  Now it
is easy to imagine that one can adjust $A$ to match the flux variance,
$\langle \delta_f^2 \rangle$ for density fields with very different
variances, $\langle \delta_\rho^2 \rangle$.  However, the skewness
depends on the variance of the density field, $\langle \delta_\rho^2
\rangle$, and on $A$ in a different way than the flux variance
$\langle \delta_f^2 \rangle$, as one can see in Figures~10-12 of
Gaztanaga \& Croft (1999).  Hence, the information in the skewness
provides an effective way of breaking the degeneracy between the
amplitude of the ionizing background and the amplitude/slope of the
matter power spectrum.

\begin{figure}[t]
\vbox{ \centerline{ \epsfig{file=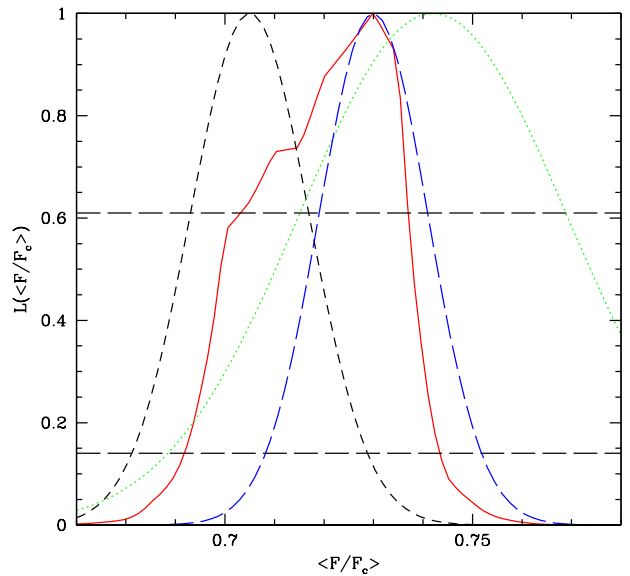,width=9.5truecm}}
  \figcaption[]{Reduced likelihood function for the
mean flux, $\langle F/F_c \rangle$, marginalized over all of the other
parameters at $z = 2.72$.  The solid red line is our constraint from
combining the flux pdf with the flux power spectrum.  The green dotted
line is the `expanded error' prior suggested by SMM03, based on local
continuum-fitting estimates of the mean transmitted flux.  The black
short-dashed line is the prior adopted by Croft et al.  (2002) based
on the PRS measurements of the mean transmitted flux.  The blue
long-dashed line is the prior adopted by Viel et al.  (2004) based on
their own local continuum-fitting estimates of the mean transmitted flux.
\label{meanf_const}}}
\end{figure}

Having demonstrated the sensitivity of the flux pdf to $\langle F/F_c
\rangle$, we constrain $\langle F/F_c \rangle$ at $z=2.72$ by
marginalizing over all of the other parameters.  The results are shown
in Figure~\ref{meanf_const}, the reduced likelihood function for
$\langle F/F_c \rangle$ using joint constraints from the flux power
spectrum and the flux pdf.  The allowed 1-$\sigma$ range is $\langle
F/F_c \rangle = 0.730 + 0.007 - 0.027$, while the allowed 2-$\sigma$
range is $\langle F/F_c \rangle = 0.730 + 0.014 - 0.038$.  The quoted
central value, $\langle F/F_c \rangle = 0.730$, corresponds to the
mean transmitted flux that maximizes the likelihood function.  The
reduced likelihood function with the current data set is hence rather
broad, only weakly constraining values of the mean transmitted flux
less than our central value.  However, we reiterate that our
constraints on the mean transmitted flux {\em do not rely on
`continuum fitting'} and are completely independent of the usual
measurements.  Next we compare our constraints with the priors adopted
on the mean transmitted flux in the Ly$\alpha$ forest adopted by other
authors.  The prior adopted by Croft et al. (2002) is $\langle F/F_c
\rangle = 0.705 \pm 0.012$, the prior of Viel et al. (2004b) is
$\langle F/F_c \rangle = 0.730 \pm 0.011$, and SMM03 consider two
priors on the mean transmitted flux: $\langle F/F_c \rangle = 0.742
\pm 0.012$, and one with expanded error bars, $\langle F/F_c \rangle =
0.742 \pm 0.027$.  Our results are thus intermediate between the Croft
et al. (2002) prior and the SMM03 prior, and do not strongly constrain
either possibility.  From Figure~\ref{meanf_const}, one can see,
however, that our constraint is narrower than the prior adopted by
SMM03.

These constraints are conservative in the sense that we assume very
little about the shape and amplitude of the linear matter power
spectrum. We can turn things around by requiring $\Lambda$CDM and
examining how tight the constraints on the mean transmitted flux
become. Specifically, we adopt a $5 \%$ prior on the slope of the
linear power spectrum, and a $10 \%$ prior on the amplitude of the
linear power spectrum, centered around a scale-invariant model with
$\sigma_8(z=0)=0.9$. The constraint on the mean transmission
(2-$\sigma$) in the Ly$\alpha$ forest becomes $\langle F/F_c \rangle =
0.724 \pm 0.012$.  This can be compared to the (2-$\sigma$) constraint
we obtain without any prior on the matter power spectrum, $\langle
F/F_c \rangle = 0.730 + 0.014 - 0.038$. The prior on the matter power
spectrum thus disfavors the very small values of the mean transmission
that are allowed without the prior.

\begin{figure}[t]
\vbox{ \centerline{ \epsfig{file=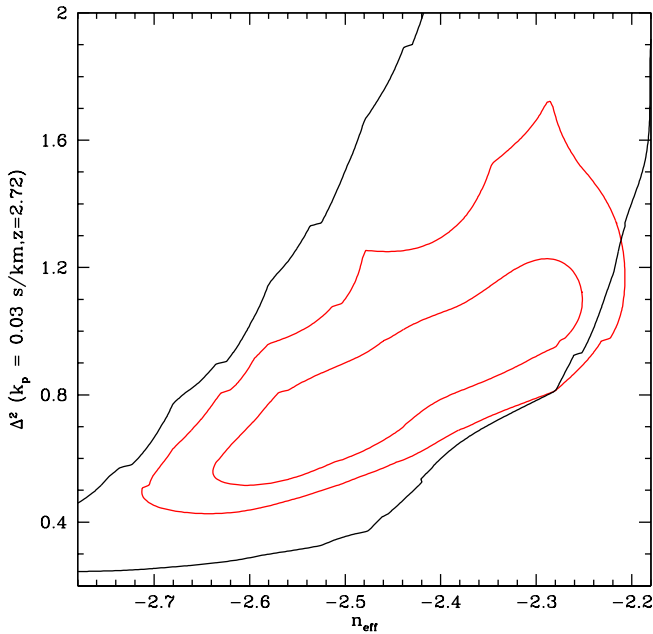,width=10.0truecm}}
  \figcaption[]{The red contours show 1- and 2-$\sigma$ constraints on
the mass power spectrum using both the probability distribution
function of $\delta_f$ and the auto spectrum, without assuming any
prior on the mean transmitted flux in the Ly$\alpha$ forest. The black
contour shows the 2-$\sigma$ constraint on the matter power spectrum
using only the auto spectrum, without enforcing any prior on $\langle
F/F_c \rangle$.   
\label{powspec_const}}}
\end{figure}

How tight do the constraints on the slope and the amplitude of the
mass power spectrum become, when assuming nothing about the mean
transmission but using the flux pdf? In Figure~\ref{powspec_const} we
show the constraint on the mass power spectrum amplitude and slope
using the information from both the flux pdf and the flux auto
spectrum. One can see that while the constraints on the mass power
spectrum amplitude and slope are still fairly weak, using the flux pdf
as well as the auto spectrum significantly tightens the
constraints. The flux pdf disfavors the high values of the mean
transmitted flux that fit the flux power spectrum with high amplitude
mass power spectrum normalizations. The information in the flux pdf
thereby helps to break the degeneracy between the mean transmitted
flux and the mass power spectrum normalization/slope by tightening
constraints on the mean transmitted flux. In the Appendix we estimate
the systematic error on our constraints resulting from the limited
resolution of our simulations.

\section{Constraints on the Slope of the Temperature-Density Relation}
\label{alpha_constrain}

\begin{figure}[t]
\vbox{ \centerline{ \epsfig{file=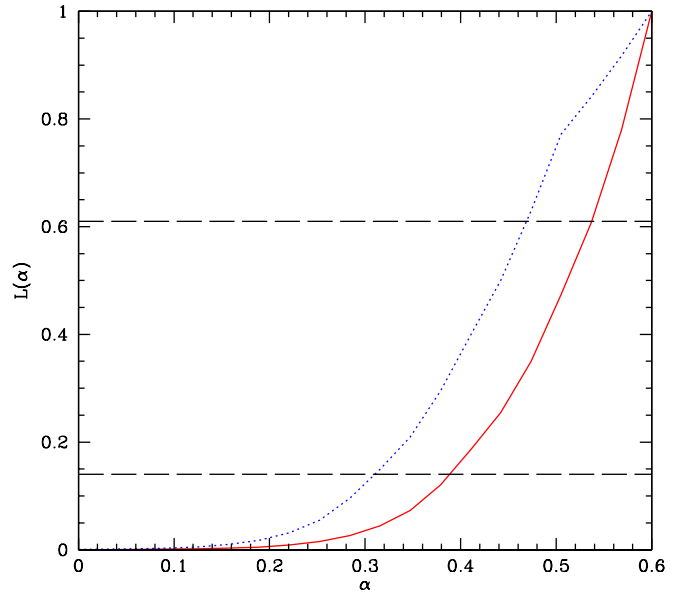,width=10.0truecm}}
  \figcaption[]{Reduced likelihood function for the slope of the
temperature density relation, $\alpha$. The red solid line shows the
likelihood function for $\alpha$ from comparing theoretical models
with the observed flux power spectrum and flux pdf at $z = 2.72$. The
blue dotted line illustrates that our results only weaken slightly if
we relax our adopted prior on the temperature at mean density. 
\label{alpha_const}}}
\end{figure}

In this section we illustrate that our measurement of the flux pdf
also yields a constraint on the slope of the temperature-density
relation, $\alpha$, at $z=2.72$.  The relation between flux and
density depends on the slope of the temperature-density relation
[Eqn.~(\ref{tau})].  We find that, when all of the other modeling
parameters are fixed, the probability of a pixel having no Ly$\alpha$
absorption becomes smaller as $\alpha$ increases (Gaztanaga \& Croft
1999).  This distinction is somewhat washed out by the smoothing we
apply to the spectra.  In spite of this we still find a constraint on
the slope of the temperature-density relation, $\alpha$.  In
Figure~\ref{alpha_const}, we show the reduced likelihood function for
the parameter $\alpha$ obtained by marginalizing over all of the other
parameters.  From the figure one can see that $\alpha \gtrsim 0.4$ is
required by our measurements at the 2-$\sigma$ level.  We checked that
this constraint does not depend significantly on our assumed prior on
the temperature at mean density, $T_0 = 300 \pm 50$ (km/s)$^2$.
Relaxing this prior, the constraint only weakens to requiring $\alpha
\gtrsim 0.32$ at 2-$\sigma$.  This constraint is surprising if HeII is
reionized at redshifts slightly larger than $z \sim 2.72$, in which
case we expect the slope of the temperature-density relation to be
close to isothermal (Hui \& Gnedin 1997).  We caution however, that we
only considered one thermal history in calculating the gas pressure
smoothing in our HPM simulations.  Desjacques \& Nusser (2004) find
that the effect of varying the amount of gas pressure smoothing on the
statistics of the Ly$\alpha$ forest is degenerate with the effect of
varying the slope of the temperature-density relation.  Our constraint
on the temperature-density relation is likely to weaken if we properly
marginalize over the amount of gas pressure smoothing.

\section{Discussion}
\label{discussion}

In this paper we have demonstrated that the information in the flux
pdf, when used in conjunction with the flux power spectrum, can
tighten constraints on the matter power spectrum. In particular, the
flux pdf of $\delta_f$ is robust to our treatment of the quasar
continuum, yet sensitive to the amplitude of the photo-ionizing
background, a sensitivity that allows us to break degeneracies between
the matter power spectrum and the amplitude of the ionizing
background. Hence, we can constrain the matter power spectrum without
relying on measurements of the mean transmitted flux which depend on
estimating the normalization of the quasar continuum.

It is appropriate to consider the prospects for using this methodology
to obtain tighter constraints on cosmology and the physics of the IGM
in the future, especially since our present constraints are still
quite weak. Observationally, the pdf of $\delta_f$ can be measured
with very high statistical precision using the SDSS (Burgess, Burles
et al., 2005, in prep.).  Theoretically, two main improvements can be
expected. First, we will soon be able to simulate the same
cosmological volume at higher resolution. This will allow us to
utilize the small-scale information obtained in the flux power
spectrum and the flux pdf and eliminate systematic errors due to poor
resolution.  Second, our pseudo-hydrodynamic approach should be
compared systematically with full hydrodynamic simulations (Gnedin \&
Hui 1998, Meiksin \& White 2001, McDonald 2004b, Viel et al. 2005b). Alternatively, it
may be feasible to run a grid of fully hydrodynamic simulations in the
near future.

There are a few other issues that would be interesting to pursue.
First, the methodology described here can be applied at a range of
redshifts to constrain the evolution of the slope of the temperature
density relation, the mean transmitted flux and the amplitude of the
ionizing background, as well as the slope and amplitude of the linear
matter power spectrum.  Our methodology may actually be most useful at
high redshift, where we did not compare our measurements with
theoretical models. At high redshift, continuum-fitting should be most
problematic since there are then zero, or very few, unabsorbed regions
of spectra with which to define the continuum.  Applying our method
over a range of redshifts would also provide constraints on the
evolution of the linear matter power spectrum with redshift, offering
a test of the extent to which the structure in the Ly$\alpha$ forest
is governed by gravitational instability (Croft et al. 2002).  Second,
the probability distribution of $\delta_f$ might be useful in
constraining primordial non-gaussianity (Gaztanaga \& Croft 1999).
Third, we can generalize our measurements to examine the pdf as a
function of smoothing scale.  Finally, the flux pdf might be useful in
constraining the impact of non-gravitational processes on the
Ly$\alpha$ forest. Gravitational instability makes definite
predictions for the relationship between the variance of the density
field and its higher order moments -- to the extent that the
Ly$\alpha$ forest is dominated by gravitational instability, these
scalings should be imprinted on the flux statistics in the Ly$\alpha$
forest (Zaldarriaga et al. 2001b, Fang \& White 2004). The good
agreement of our measured flux pdf and simulations should therefore
imply a constraint on feedback processes.

\section*{{Acknowledgements}}

AL and LH were supported in part by the Outstanding Junior
Investigator Award from the Department of Energy, and AST-0098437
grant from the NSF. We thank Kristen Burgess, Scott Burles and Josh
Frieman for useful conversations. We acknowledge parallel computing
support from Los Alamos National Laboratory's Institutional Computing
Initiative. This research was supported by the DOE, under contract
W-7405-ENG-36.

\section*{Appendix}
\label{appendix}

The primary purpose of this section is to examine the sensitivity of
our results to limitations in our simulation box size and
resolution. The $\Lambda$CDM simulations in this paper use $512^3$
baryons and $512^3$ dark matter particles, with $512^3$ mesh points,
in a $40$ Mpc/$h$ box.  The simulations were run with a fast, parallel
N-body code, MC$^2$ ({\bf M}esh-based {\bf C}osmology {\bf C}ode).
MC$^2$ was recently extensively tested against five other
state-of-the-art codes (Heitmann et al.  2004).  A detailed
description of the code will be given in Habib et al.  (2005) (in
preparation), including a description of our implementation of HPM,
the pseudo-hydrodynamic method of Gnedin \& Hui (1998).  Our
implementation differs from that of Gnedin \& Hui (1998) in that we
follow separately two particle species: baryonic particles that
experience gas forces as well as gravity, and dark matter particles
that interact only gravitationally. We adopt the input transfer
function of Hu and Sugiyama (1996) with cosmological parameters of
$\Omega_m=0.3, \Omega_\Lambda=0.7, \Omega_b h^{2}=0.02$. Simulations
with differing spectral indices, $n$, are each normalized so as to
have the same amplitude, $\Delta^2(\tilde{k}_p,z)$, on the scale
$\tilde{k}_p=0.008$ s/km at a redshift of $z=3$.  The output with
$n=1.0$ has a normalization given by $\sigma_8 (z=0) = 0.84$.
Finally, we need to specify the thermal history of the IGM in order to
compute the gas pressure term in our HPM simulation. We adopt a
similar thermal history to that adopted in McDonald \&
Miralda-Escud\'e (2001). Specifically, we assume that $T_0$ is
$25,000$ K and that the slope of the temperature-density relation is
$\alpha = 0$ at a reionization redshift of $z_{\rm reion} = 10$, while
$T_0 = 18,000$ K and $\alpha = 0.3$ at $z = 4$. We linearly
interpolate between these values to specify the gas pressure at
intermediate redshifts, and use the $z = 4$ values at lower
redshifts. While the true thermal history of the IGM may be rather
different than this simple model, we find that the flux power spectrum
on the large scales considered in this paper, $k \lesssim 0.02$ s/km,
are relatively insensitive to the thermal history. More systematic
tests of the dependence of our results on thermal history are clearly
warranted, however.

\begin{figure}[t]
\vbox{ \centerline{ \epsfig{file=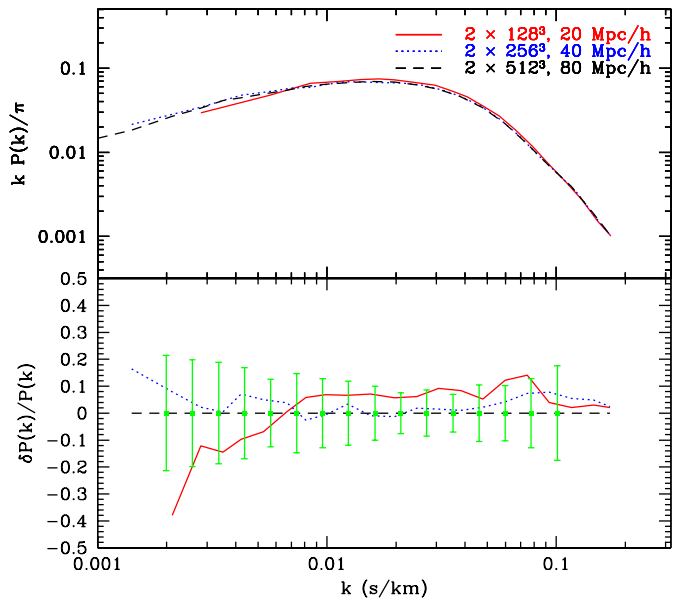,width=10.0truecm}}
  \figcaption[]{Convergence test of the flux power spectrum with box
size at $z=3.0$. In the top panel, we show the flux auto spectrum in
$20,40$ and $80$ Mpc/$h$ simulation boxes. Each model has the same
resolution, corresponding to $2 \times 128^3$, $2 \times 256^3$, and
$2 \times 512^3$ particles, respectively. The model is described by
$(a, n, T_0, \alpha, \langle F/F_c \rangle)= (0.2480, 1.00, 300$
(km/s)$^2$, $0.4, 0.684).$ The bottom panel shows the fractional
difference between the flux power spectrum in each of the $20$ and
$40$ Mpc/$h$ simulation boxes with that in the $80$ Mpc/$h$ simulation
box. The green points show the size of the (2-$\sigma$) fractional
error bars from Croft et al. (2002) at $z=2.72$.   
\label{boxsize}}}
\end{figure}

In Figure~\ref{boxsize}, we show flux power spectrum measurements at
$z=3$ for simulations with box sizes of $20, 40$ and $80$ Mpc/$h$.
Each simulation has a resolution equivalent to that of a $2 \times
512^3$ particle, $80$ Mpc/$h$ simulation. For the purpose of testing
the resolution and box size of our simulations, we generate flux
fields at $z=3.0$ for a model with $(a, n, k_f, T_0, \alpha, \langle
F/F_c \rangle) = (0.2480, 1.0, 300$ (km/s)$^2$, $0.4, 0.684).$ For
this model, the box size convergence test shown in the figure
indicates that any difference in the flux power spectrum between the
$40$ Mpc/$h$ simulation and the $80$ Mpc/$h$ is at most comparable to
the (1-$\sigma$) statistical error on the Croft et al. (2002)
measurement, and always less than the 2-$\sigma$ statistical errors
shown in the figure. Furthermore, much of the difference between the
$40$ Mpc/$h$ simulation and the $80$ Mpc/$h$ simulation is likely
random, as opposed to systematic, owing simply to differences in the
random initial conditions adopted in the two simulations.  Therefore,
a $40$ Mpc/$h$ simulation box is adequate to achieve convergence at
the level of the Croft et al. (2002) error bars.

\begin{figure}[t]
\vbox{ \centerline{ \epsfig{file=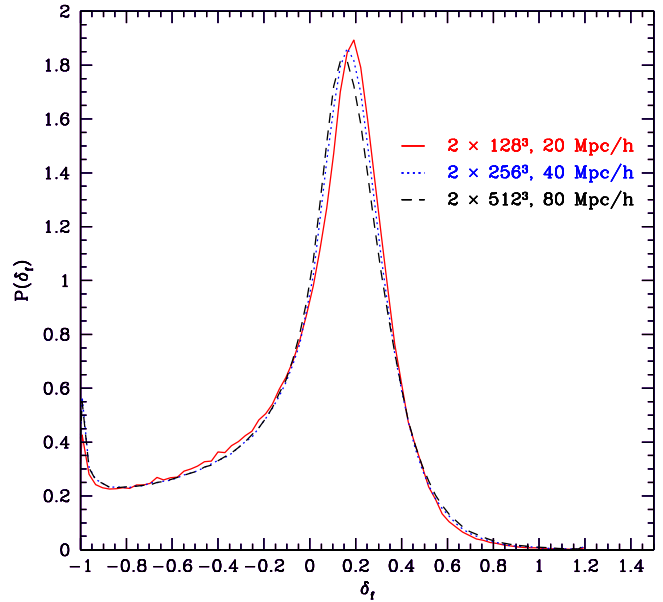,width=10.0truecm}}
  \figcaption[]{Convergence test of the flux pdf with box size at
$z=3.0$. The simulation flux fields have been smoothed with filters of
$R=500$ km/s and $r=30$ km/s. The flux pdf is shown in simulation
boxes of size $20,40$ and $80$ Mpc/$h$, each with the same resolution.
\label{pdf_boxsize}}}
\end{figure}

In Figure~\ref{pdf_boxsize}, we show the equivalent convergence test
for the flux pdf.  In particular, it is important to confirm that our
results converge with box size, in spite of the size we choose for our
large-radius filter, $R=500$ km/s, which is roughly 1/8th the size of
the $40$ Mpc/$h$ simulation box. The figure illustrates that any
difference between the flux pdf in a $40$ Mpc/$h$ simulation box and
that in a $80$ Mpc/$h$ simulation box is very small.

\begin{figure}[t]
\vbox{ \centerline{ \epsfig{file=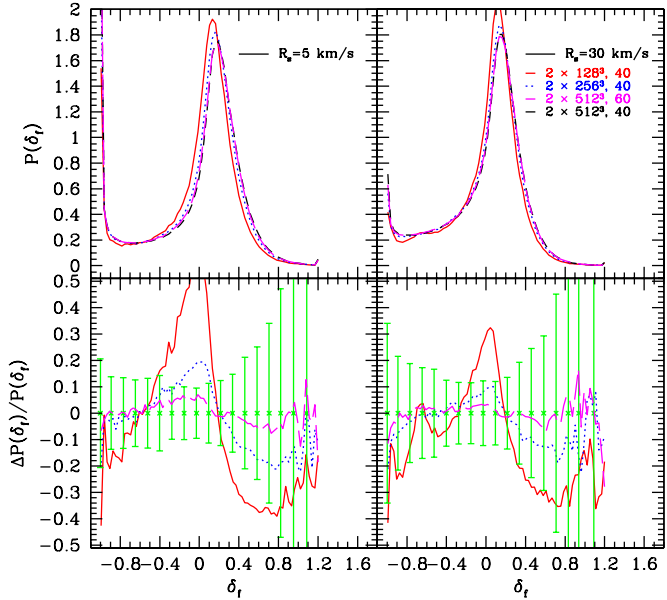,width=10.0truecm}}
  \caption[]{  \label{pdf_resolution}      
    Convergence test of the flux pdf with resolution at $z=3.0$. We
show the flux pdf in a $40$ Mpc/$h$ simulation box for simulations
with $2 \times 128^3$, $2 \times 256^3$, and $2 \times 512^3$
particles, as well as the flux pdf in a $60$ Mpc/$h$ simulation with
$2 \times 512^3$ particles. The panels on the left side show the flux
pdf for a small-scale smoothing of $r=5$ km/s and the panels on the
right hand side show the flux pdf for a small-scale smoothing of
$r=30$ km/s.  In each case the flux fields were smoothed with the same
large radius filter of $R=500$ km/s.  The top panels show the
measurements themselves, while the bottom panels show the fractional
difference between the lower resolution measurements and the high
resolution measurement.  The green points represent the approximate
(1-$\sigma$) fractional error from the present data sample.}}
\end{figure}

Another important aspect of our modeling is the choice of the
smoothing scale, $r$, at which we smooth the data and simulated
spectra in order to measure the pdf of $\delta_f$. We aim to choose
this scale to be sufficiently large that we can guarantee the
convergence of the flux pdf, when the flux fields are smoothed on the
scale $r$, with increasing resolution. To test the convergence of the
flux pdf with resolution we have generated the flux pdf assuming a
range of different small-scale smoothings, $r$, in simulations of
varying resolution. (The varying resolution simulations are described
in more detail below.) In Figure~\ref{pdf_resolution} we show the
convergence of our simulation measurements with increasing resolution
for small-scale smoothings of $r=5$ km/s and $r=30$ km/s. As expected,
the convergence is better in the case of the $r=30$ km/s filter than
in the case of the $r=5$ km/s filter. In the case of the $r=30$ km/s
filter it is clear that the convergence of our results is adequate for
our present data sample. The convergence also appears surprisingly
good in the case of the $r=5$ km/s filter: the difference between the
$256^3$, $40$ Mpc/$h$ and the $512^3$, $40$ Mpc/$h$ simulation is only
slightly larger than the statistical error bars. However, in our
present analysis we conservatively adopt the $r=30$ km/s filter.

\begin{figure}[t]
\vbox{ \centerline{ \epsfig{file=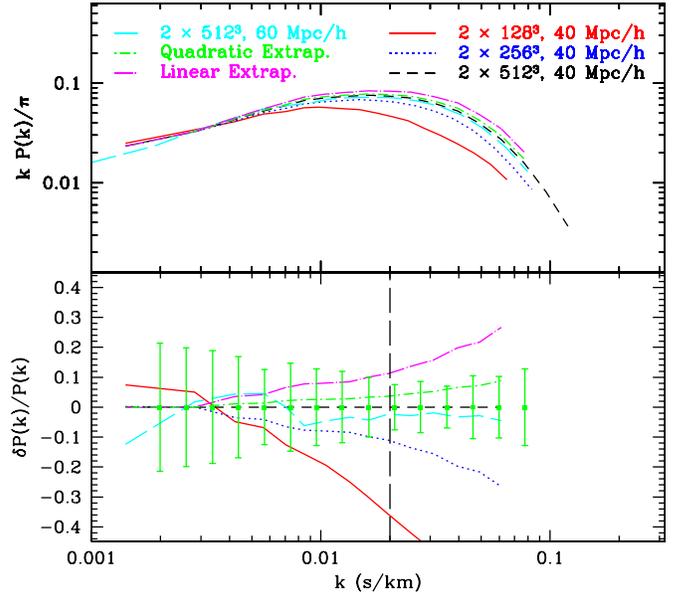,width=10.0truecm}}
  \figcaption[]{Convergence test of the flux auto spectrum with
resolution at $z=3.0$. We show the flux auto spectrum in a $40$
Mpc/$h$ simulation box for each of $2 \times 128^3, 2 \times 256^3$,
and $2 \times 512^3$ particles. We also show the flux auto spectrum in
a $2 \times 512^3$ particle, $60$ Mpc/$h$ simulation box, and two
extrapolations of the flux auto spectrum to infinite resolution. In
one extrapolation we assume that the flux power in each $k$-mode
converges linearly with the mesh size, and in the other extrapolation
we assume that the flux power converges quadratically with the mesh
size. The top panel shows the flux power, while the bottom panel shows
the fractional difference with the $512^3$, $40$ Mpc/$h$
simulation. The green points show the size of the (2-$\sigma$)
fractional error bars from Croft et al. (2002) at $z=2.72$. In our
present analysis, we only include measurements on scales larger than
that indicated by the vertical dashed line.    
\label{resolution}}}
\end{figure}

Next, we examine the convergence of the flux power spectrum with
resolution. To perform this test, we measure the flux power spectrum
for several simulations of the same box size, but differing particle
number. Specifically, we measure the flux power spectrum in $40$
Mpc/$h$ simulations with each of $2 \times 128^3, 2 \times 256^3$ and
$2 \times 512^3$ particles. Each of the $40$ Mpc/$h$ simulations are
generated with similar initial conditions: the initial conditions for,
e.g., the $2 \times 256^3$ simulation are generated by averaging the
initial conditions for the $2 \times 512^3$ particle simulation over
every eight particles. We also generate a $2 \times 512^3$ particle,
$60$ Mpc/$h$ simulation to further illustrate the convergence of our
results with varying mesh size. This simulation is generated with
different initial conditions than that of the $40$ Mpc/$h$
simulations, but Figure~\ref{boxsize} demonstrates that the resulting
random differences between the flux power spectra should be small for
the $k$-modes of interest, $k \gtrsim 0.01$ s/km.  Finally, we
extrapolate our results to the case of infinite resolution using
Richardson extrapolation. We make this extrapolation in two
ways. First we assume that the convergence of the flux power spectrum
for a given $k$-mode is linear in the mesh size, and second we assume
that the convergence is quadratic in the mesh size.  The results of
our resolution test are shown in Figure~\ref{resolution}.
Interestingly, the convergence appears better in the flux pdf than in
the auto spectrum, although this is partly due to the large error bars
on the pdf measurement; our pdf estimate comes from a smaller data
sample than the sample Croft et al. (2002) use to estimate the flux
power spectrum.  At any rate, the test shows rather large systematic
differences between the lowest resolution simulation, with $2 \times
128^3$ particle and the $2 \times 512^3$ particle, $40$ Mpc/$h$
simulation.  These differences are expected since the lowest
resolution simulations do not resolve the gas pressure smoothing
scale. The measurements made using the $2 \times 256^3$, $40$ Mpc/$h$
simulation and the $2 \times 512^3$, $60$ Mpc/$h$ simulation are
substantially closer to those from the $2 \times 512^3$, $40$ Mpc/$h$
simulation.

How close is the flux power spectrum in our $512^3$, $40$ Mpc/$h$
simulation to what we would measure from an infinite resolution
simulation? This depends, of course, on how we extrapolate our
measurements to infinite resolution.  If the convergence is as good as
quadratic in the mesh size, then our present simulations are adequate:
any systematic difference between the $2 \times 512^3$, $40$ Mpc/$h$
simulation and the quadratic extrapolation to perfect resolution is
small compared to the statistical error bars on the Croft et
al. (2002) measurement. However, if the convergence is only linear in
the mesh size, then the figure illustrates that, even limiting our
analysis to scales of $k \lesssim 0.02$ s/km, we are making some
non-negligible systematic error. Specifically, assuming linear
convergence, our $2 \times 512^3$, $40$ Mpc/h flux power spectrum is
good to $\sim 10\%$. While this convergence appears to be quite
acceptable, it is in fact comparable to the 2-$\sigma$ statistical
error on the Croft et al. (2002) measurement at $z=2.72$. It is
important to decide which is a more accurate description of the
convergence of our results with mesh size: linear or quadratic
convergence? At present, the best we can do to answer this question is
to ask how well we can predict the results of the $2 \times 512^3, 40$
Mpc/$h$ simulation by extrapolating from the lower resolution
simulations. On this basis, we find that neither linear nor quadratic
convergence is a good description over the full range of $k$-modes we
consider. At the $k$-modes that our most important for our convergence
study, $k \sim 0.02$ s/km, by making the (worse-case) assumption that
linear convergence is an accurate description, we can check how large
a systematic error we are possibly making due to imperfect simulation
resolution.

\begin{figure}[t]
\vbox{ \centerline{ \epsfig{file=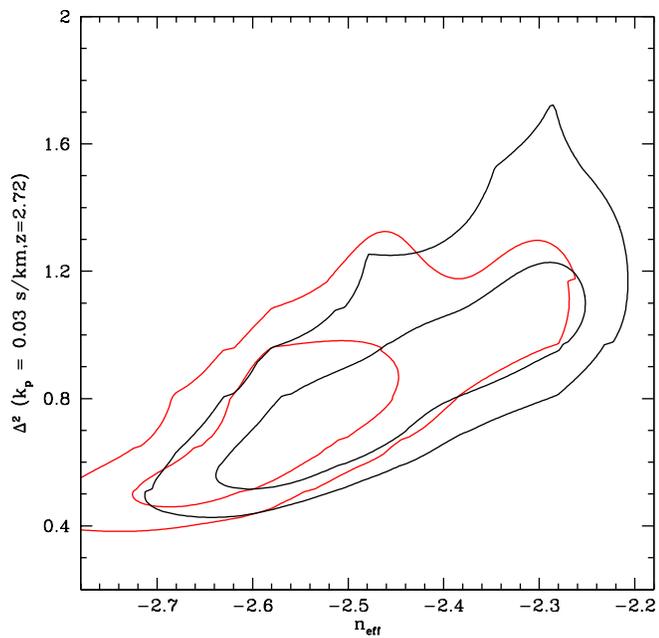,width=10.0truecm}}
   \figcaption[]{An estimate of the systematic error in
our constraints on the linear matter power spectrum owing to the
imperfect resolution of our numerical simulations. The black contours
indicate 1- and 2-$\sigma$ constraints on the slope and amplitude of
the linear matter power spectrum. The red contours are estimates of
the same constraints we expect given infinite resolution simulations. 
\label{system_resolution}}}
\end{figure}

In order to gauge the importance of any systematic error from the
limited resolution of our simulations we do the following check. We
assume that the convergence of our flux power spectrum measurement is
linear in the mesh size, and derive a correction factor at each
$k$-mode by dividing the linearly-extrapolated flux power by that in
our $2 \times 512^3$ particle, $40$ Mpc/$h$ simulation box (see also
McDonald et al.  2004b). Assuming that this correction factor is
model-independent, we then multiply each model in our grid by this
correction and examine how much our constraints on the slope and
amplitude of the matter power spectrum change. This is clearly an
imperfect procedure: the relevant correction factor is likely
model-dependent and our assumption of linear convergence is not
strongly established. In spite of these uncertainties, this test gives
some indication of how much our results might change with improved
resolution.  This is shown in Figure~\ref{system_resolution}, where
the black contours show our current constraints, and the red contours
indicate our estimate of the same constraints given simulations of
infinite resolution. From this plot, the main effect of our limited
resolution appears to be to a systematic shift in the inferred slope
of the matter power spectrum towards steeper slopes by $\sim 0.1$. The
`resolution corrected' contours are also less elongated towards large
matter power spectrum amplitudes. From this test, the effect of
limited resolution appears to be most degenerate with the slope of the
matter power spectrum (see also SMM03). While our present constraints
are weak, it is clear that higher resolution simulations will be
necessary to obtain more precise constraints in the future.

\clearpage

\begin{deluxetable}{ccccccc}
\tablecaption{Quasar Spectra and Wavelength Range used in the PDF
Measurement 
\label{lambda_use}}
\tablehead{
\colhead{Quasar Spectrum} & \colhead{$z_{\rm em}$} & 
\colhead{$\lambda_{\rm low}(\AA)$} &
\colhead{$\lambda_{\rm high}(\AA)$}} 
\startdata
Q2343+123 & 2.52        & 3922.0  & 3926.6\\
          &             & 3929.3  & 3966.5\\
          &             & 3970.5  & 3996.3\\
          &             & 4000.5  & 4053.2\\
          &             & 4060.6  & 4077.\\
          &             & 4097.   & 4108.8\\
\\
Q1442+293 & 2.67        & 3764.4  & 4089.5\\
          &             & 4095.1  & 4100.2\\
          &             & 4105.5  & 4156.2\\
          &             & 4204.5  & 4250.0\\
          &             & 4336.8  & 4375.5\\
          &             & 4379.6  & 4409.28\\
\\
Q1107+485 & 3.00        & 4233.0   & 4261.3\\
          &             & 4262.7   & 4306.1\\
          &             & 4308.8   & 4364.7\\
          &             & 4365.4   & 4408.5\\
          &             & 4409.5   & 4434.5\\
          &             & 4437.5   & 4444.8\\
          &             & 4446.5   & 4485.7\\
          &             & 4491.9   & 4534.8\\
          &             & 4541.9   & 4549.4\\
          &             & 4550.9   & 4656.3\\
          &             & 4660.3   & 4670.1\\
          &             & 4678.9   & 4698.2\\
          &             & 4701.3   & 4741.3\\
          &             & 4744.9   & 4797.9\\
\\
Q1425+604 & 3.20        & 4314.0   & 4320.7\\
          &             & 4321.9  & 4373.0\\
          &             & 4375.5  & 4462.5\\
          &             & 4467.5  & 4486.0\\
          &             & 4490.6  & 4492.3\\
          &             & 4502.7  & 4546.0\\
          &             & 4571.9  & 4591.3\\
          &             & 4593.8  & 4599.0\\
          &             & 4603.5  & 4606.6\\
          &             & 4753.9  & 4821.1\\
          &             & 4829.0  & 4858.1\\
          &             & 4862.3  & 4905.4\\
          &             & 4910.0  & 4916.3\\
          &             & 4918.0  & 4980.5\\
          &             & 4996.9  & 5032.0\\
\\
Q1422+230 & 3.62        & 4738.1   & 4786.4\\
          &             & 4818.0   & 4884.5\\
          &             & 4885.7   & 4890.6\\
          &             & 4892.2   & 4980.7\\
          &             & 4982.2   & 5002.3\\
          &             & 5003.7   & 5224.3\\
          &             & 5225.8   & 5287.3\\
          &             & 5288.5   & 5364.6\\
          &             & 5369.0   & 5472.4\\
          &             & 5477.0   & 5523.0\\
\\
\enddata
\tablecomments{Table continued on the next page. The quasar spectra used in the pdf measurement. The wavelength
range specified indicates the pixels that are used in the analysis, other
data are excluded due to the expected existence of metal lines, damped 
lyman-alpha systems, or spurious pixels. See also Rauch et al. (1997)
and McDonald et al. (2000).}
\end{deluxetable}

\newpage

\begin{deluxetable}{ccccccc}
\tablecaption{Quasar Spectra and Wavelength Range used in the PDF
Measurement (continued)
\label{lambda_use_cont}}
\tablehead{ \colhead{Quasar Spectrum} & \colhead{$z_{\rm em}$} &
  \colhead{$\lambda_{\rm low}(\AA)$} & \colhead{$\lambda_{\rm
      high}(\AA)$}} \startdata
Q000-262  & 4.11        & 5600.0   & 5637.0\\
&             & 5639.8  & 5704.4\\
&             & 5706.4  & 5715.2\\
&             & 5718.2  & 5724.9\\
&             & 5727.2  & 5813.5\\
&             & 5815.3  & 5859.7\\
&             & 5860.8  & 5953.5\\
&             & 5956.4  & 6092.0\\
\\
Q2237-061 & 4.55        & 5692.0  & 5703.8\\
&             & 5705.0  & 5747.4\\
&             & 5749.7  & 5774.4\\
&             & 5774.8  & 5803.3\\
&             & 5805.7  & 5815.3\\
&             & 5817.6  & 5830.4\\
&             & 5831.5  & 5876.9\\
&             & 5878.1  & 5891.1\\
&             & 5892.0  & 5897.0\\
&             & 5898.0  & 5967.0\\
&             & 5967.2  & 5974.8\\
&             & 5975.6  & 6020.1\\
&             & 6020.8  & 6034.6\\
&             & 6036.5  & 6042.7\\
&             & 6055.5  & 6063.4\\
&             & 6068.9  & 6100.0\\
&             & 6250.0  & 6251.6\\
&             & 6255.4  & 6282.3\\
&             & 6290.3  & 6314.5\\
&             & 6315.5  & 6330.6\\
&             & 6332.3  & 6402.3\\
&             & 6405.8  & 6482.3\\
&             & 6485.8  & 6502.0\\
&             & 6503.5  & 6549.1\\
&             & 6550.0  & 6582.1\\
&             & 6582.2  & 6600.0\\
\enddata \tablecomments{Table continued from the previous page. The
  quasar spectra used in the pdf measurement. The wavelength range
  specified indicates the pixels that are used in the analysis, other
  data are excluded due to the expected existence of metal lines,
  damped lyman-alpha systems or spurious pixels. See also 
  Rauch et al. (1997) and McDonald et al.
  (2000).}
\end{deluxetable}

\begin{table*}[ht]
\centering
\caption[pdftable]{\label{pdftable}
The probability distribution of $\delta_f$ at $\langle z \rangle = 2.72$.}
\begin{tabular}{cccccccc}
\hline &\\
Bin No. & $\delta_f$ & $P(\delta_f)$ & $\sigma^2_P(\delta_f)$ 
& Bin No. & $\delta_f$ & $P(\delta_f)$ & $\sigma^2_P(\delta_f)$ \\
\hline &\\
      1 & -0.9983E+00 &  0.3483E+00 &  0.1406E-01 &     19 &  0.9415E-01 &  0.1964E+01 &  0.5558E-01\\
      2 & -0.9596E+00 &  0.3390E+00 &  0.4917E-02 &     20 &  0.1526E+00 &  0.1872E+01 &  0.3540E-01\\
      3 & -0.8913E+00 &  0.1718E+00 &  0.1371E-02 &     21 &  0.2142E+00 &  0.1714E+01 &  0.5379E-01\\
      4 & -0.8317E+00 &  0.2090E+00 &  0.1600E-02 &     22 &  0.2739E+00 &  0.1235E+01 &  0.4550E-01\\
      5 & -0.7706E+00 &  0.1858E+00 &  0.1213E-02 &     23 &  0.3366E+00 &  0.6548E+00 &  0.1588E-01\\
      6 & -0.7013E+00 &  0.1486E+00 &  0.1105E-02 &     24 &  0.3985E+00 &  0.4273E+00 &  0.8478E-02\\
      7 & -0.6450E+00 &  0.1765E+00 &  0.9211E-03 &     25 &  0.4584E+00 &  0.3530E+00 &  0.7458E-02\\
      8 & -0.5853E+00 &  0.2276E+00 &  0.1547E-02 &     26 &  0.5206E+00 &  0.1533E+00 &  0.1817E-02\\
      9 & -0.5259E+00 &  0.2322E+00 &  0.1491E-02 &     27 &  0.5837E+00 &  0.1486E+00 &  0.1890E-02\\
     10 & -0.4593E+00 &  0.2647E+00 &  0.1542E-02 &     28 &  0.6443E+00 &  0.1207E+00 &  0.1695E-02\\
     11 & -0.4016E+00 &  0.3669E+00 &  0.2094E-02 &     29 &  0.7081E+00 &  0.5109E-01 &  0.5303E-03\\
     12 & -0.3379E+00 &  0.3715E+00 &  0.2788E-02 &     30 &  0.7693E+00 &  0.6966E-01 &  0.7736E-03\\
     13 & -0.2766E+00 &  0.3437E+00 &  0.1596E-02 &     31 &  0.8312E+00 &  0.5109E-01 &  0.7147E-03\\
     14 & -0.2161E+00 &  0.3622E+00 &  0.2797E-02 &     32 &  0.8806E+00 &  0.1393E-01 &  0.6471E-04\\
     15 & -0.1519E+00 &  0.5387E+00 &  0.3878E-02 &     33 &  0.9382E+00 &  0.2322E-01 &  0.2808E-03\\
     16 & -0.9288E-01 &  0.6966E+00 &  0.7684E-02 &     34 &  0.1014E+01 &  0.4644E-02 &  0.3863E-04\\
     17 & -0.2913E-01 &  0.8545E+00 &  0.1101E-01 &     35 &  0.1091E+01 &  0.4644E-02 &  0.2454E-04\\
     18 &  0.3326E-01 &  0.1547E+01 &  0.7748E-01 &     36 &  0.1130E+01 &  0.4644E-02 &  0.9799E-04\\    \hline &&\\
\tablecomments{The one point probability distribution of flux as a
function of $\delta_f$. The first four columns are respectively the
bin number, the average $\delta_f$ in the bin, the average pdf in the bin, and
a jackknife estimate of the error in the pdf. The fifth through eighth
columns are the same as the first four columns for higher flux
bins.} 
\end{tabular} 
\end{table*}

\begin{table*}[ht]
\centering
\caption[pdftable]{\label{pdf_bin1}
The probability distribution of $\delta_f$ at $\langle z \rangle = 2.26$.}
\begin{tabular}{cccccccc}
\hline &\\
Bin No. & $\delta_f$ & $P(\delta_f)$ & $\sigma^2_P(\delta_f)$ 
& Bin No. & $\delta_f$ & $P(\delta_f)$ & $\sigma^2_P(\delta_f)$ \\
\hline &\\
      1 & -0.1015E+01 &  0.1382E+00 &  0.5478E-02 &     19 &  0.8905E-01 &  0.2683E+01 &  0.1241E+00\\
      2 & -0.9534E+00 &  0.1219E+00 &  0.2655E-02 &     20 &  0.1557E+00 &  0.1894E+01 &  0.5844E-01\\
      3 & -0.8912E+00 &  0.2032E+00 &  0.8366E-02 &     21 &  0.2122E+00 &  0.1528E+01 &  0.8463E-01\\
      4 & -0.8349E+00 &  0.1138E+00 &  0.1236E-02 &     22 &  0.2716E+00 &  0.8292E+00 &  0.3612E-01\\
      5 & -0.7647E+00 &  0.1382E+00 &  0.1511E-02 &     23 &  0.3373E+00 &  0.2764E+00 &  0.9017E-02\\
      6 & -0.7096E+00 &  0.1382E+00 &  0.1374E-02 &     24 &  0.3928E+00 &  0.1707E+00 &  0.8643E-02\\
      7 & -0.6471E+00 &  0.1301E+00 &  0.1234E-02 &     25 &  0.4600E+00 &  0.9755E-01 &  0.3620E-02\\
      8 & -0.5816E+00 &  0.1301E+00 &  0.1092E-02 &     26 &  0.5145E+00 &  0.1219E+00 &  0.8451E-02\\
      9 & -0.5251E+00 &  0.1301E+00 &  0.1057E-02 &     27 &  0.5767E+00 &  0.9755E-01 &  0.4756E-02\\
     10 & -0.4579E+00 &  0.1707E+00 &  0.1856E-02 &     28 &  0.6210E+00 &  0.8129E-02 &  0.6836E-04\\
     11 & -0.4060E+00 &  0.2114E+00 &  0.1800E-02 &     29 &  0.0000E+00 &  0.0000E+00 &  0.0000E+00\\
     12 & -0.3429E+00 &  0.1707E+00 &  0.1859E-02 &     30 &  0.0000E+00 &  0.0000E+00 &  0.0000E+00\\
     13 & -0.2701E+00 &  0.2520E+00 &  0.3739E-02 &     31 &  0.0000E+00 &  0.0000E+00 &  0.0000E+00\\
     14 & -0.2121E+00 &  0.3821E+00 &  0.5326E-02 &     32 &  0.0000E+00 &  0.0000E+00 &  0.0000E+00\\
     15 & -0.1502E+00 &  0.4796E+00 &  0.4609E-02 &     33 &  0.0000E+00 &  0.0000E+00 &  0.0000E+00\\
     16 & -0.9019E-01 &  0.7560E+00 &  0.2314E-01 &     34 &  0.0000E+00 &  0.0000E+00 &  0.0000E+00\\
     17 & -0.2765E-01 &  0.1366E+01 &  0.5216E-01 &     35 &  0.0000E+00 &  0.0000E+00 &  0.0000E+00\\
     18 &  0.3238E-01 &  0.3512E+01 &  0.3496E+00 &     36 &  0.0000E+00 &  0.0000E+00 &  0.0000E+00\\      
&&\\
\tablecomments{The same as Table~(\ref{pdftable}), except for the redshift bin with
$\langle z \rangle = 2.26$ (see Table~\ref{zbins}).} 
\end{tabular} 
\end{table*}

\begin{table*}[ht]
\centering
\caption[pdftable]{\label{pdf_bin3}
The probability distribution of $\delta_f$ at $\langle z \rangle = 3.28$.}
\begin{tabular}{cccccccc}
\hline &\\
Bin No. & $\delta_f$ & $P(\delta_f)$ & $\sigma^2_P(\delta_f)$ 
& Bin No. & $\delta_f$ & $P(\delta_f)$ & $\sigma^2_P(\delta_f)$ \\
\hline &\\
      1 & -0.9947E+00 &  0.2902E+00 &  0.1450E-01 &     19 &  0.9358E-01 &  0.1049E+01 &  0.2613E-01\\
      2 & -0.9543E+00 &  0.4551E+00 &  0.8468E-02 &     20 &  0.1538E+00 &  0.1194E+01 &  0.2766E-01\\
      3 & -0.8906E+00 &  0.4353E+00 &  0.2913E-02 &     21 &  0.2154E+00 &  0.1372E+01 &  0.5276E-01\\
      4 & -0.8287E+00 &  0.3232E+00 &  0.5159E-02 &     22 &  0.2761E+00 &  0.1200E+01 &  0.1899E-01\\
      5 & -0.7692E+00 &  0.3034E+00 &  0.2115E-02 &     23 &  0.3388E+00 &  0.9695E+00 &  0.1769E-01\\
      6 & -0.7084E+00 &  0.3100E+00 &  0.2483E-02 &     24 &  0.3984E+00 &  0.7584E+00 &  0.1251E-01\\
      7 & -0.6432E+00 &  0.3166E+00 &  0.3222E-02 &     25 &  0.4634E+00 &  0.6529E+00 &  0.1901E-01\\
      8 & -0.5831E+00 &  0.2704E+00 &  0.1906E-02 &     26 &  0.5220E+00 &  0.4221E+00 &  0.4633E-02\\
      9 & -0.5225E+00 &  0.2770E+00 &  0.2094E-02 &     27 &  0.5860E+00 &  0.2770E+00 &  0.4519E-02\\
     10 & -0.4599E+00 &  0.3363E+00 &  0.3365E-02 &     28 &  0.6418E+00 &  0.3297E+00 &  0.6822E-02\\
     11 & -0.4017E+00 &  0.3891E+00 &  0.2566E-02 &     29 &  0.7016E+00 &  0.1451E+00 &  0.3897E-02\\
     12 & -0.3419E+00 &  0.4023E+00 &  0.2653E-02 &     30 &  0.7641E+00 &  0.7914E-01 &  0.1172E-02\\
     13 & -0.2799E+00 &  0.4814E+00 &  0.4446E-02 &     31 &  0.8174E+00 &  0.1319E-01 &  0.8699E-04\\
     14 & -0.2151E+00 &  0.4155E+00 &  0.4118E-02 &     32 &  0.9020E+00 &  0.3957E-01 &  0.5024E-03\\
     15 & -0.1488E+00 &  0.4814E+00 &  0.3175E-02 &     33 &  0.9599E+00 &  0.6595E-01 &  0.1799E-02\\
     16 & -0.9270E-01 &  0.6265E+00 &  0.4565E-02 &     34 &  0.1018E+01 &  0.1978E-01 &  0.3634E-03\\
     17 & -0.2955E-01 &  0.6925E+00 &  0.6691E-02 &     35 &  0.1087E+01 &  0.3957E-01 &  0.2377E-03\\
     18 &  0.3172E-01 &  0.7848E+00 &  0.1571E-01 &     36 &  0.1140E+01 &  0.3297E-01 &  0.1672E-02\\      
&&\\
\tablecomments{The same as Table~(\ref{pdftable}), except for the redshift bin with
$\langle z \rangle = 3.28$ (see Table~\ref{zbins}).} 
\end{tabular} 
\end{table*}

\begin{table*}[ht]
\centering
\caption[pdftable]{\label{pdf_bin4}
The probability distribution of $\delta_f$ at $\langle z \rangle = 3.99$.}
\begin{tabular}{cccccccc}
\hline &\\
Bin No. & $\delta_f$ & $P(\delta_f)$ & $\sigma^2_P(\delta_f)$ 
& Bin No. & $\delta_f$ & $P(\delta_f)$ & $\sigma^2_P(\delta_f)$ \\
\hline &\\
      1 & -0.1016E+01 &  0.7323E+00 &  0.4294E-01 &     19 &  0.9680E-01 &  0.5300E+00 &  0.6503E-02\\
      2 & -0.9549E+00 &  0.7253E+00 &  0.1136E-01 &     20 &  0.1511E+00 &  0.5789E+00 &  0.5608E-02\\
      3 & -0.8961E+00 &  0.4394E+00 &  0.3914E-02 &     21 &  0.2174E+00 &  0.5789E+00 &  0.7551E-02\\
      4 & -0.8317E+00 &  0.4882E+00 &  0.5043E-02 &     22 &  0.2751E+00 &  0.5440E+00 &  0.7219E-02\\
      5 & -0.7688E+00 &  0.4673E+00 &  0.3830E-02 &     23 &  0.3359E+00 &  0.5161E+00 &  0.3712E-02\\
      6 & -0.7093E+00 &  0.4673E+00 &  0.4368E-02 &     24 &  0.3970E+00 &  0.5370E+00 &  0.3745E-02\\
      7 & -0.6479E+00 &  0.4254E+00 &  0.3701E-02 &     25 &  0.4651E+00 &  0.4882E+00 &  0.3472E-02\\
      8 & -0.5825E+00 &  0.3906E+00 &  0.4653E-02 &     26 &  0.5239E+00 &  0.4533E+00 &  0.9987E-02\\
      9 & -0.5201E+00 &  0.3906E+00 &  0.4975E-02 &     27 &  0.5851E+00 &  0.3696E+00 &  0.2578E-02\\
     10 & -0.4614E+00 &  0.3627E+00 &  0.3310E-02 &     28 &  0.6461E+00 &  0.5021E+00 &  0.8142E-02\\
     11 & -0.3998E+00 &  0.4603E+00 &  0.3210E-02 &     29 &  0.7059E+00 &  0.4254E+00 &  0.2967E-02\\
     12 & -0.3389E+00 &  0.4324E+00 &  0.5804E-02 &     30 &  0.7680E+00 &  0.3975E+00 &  0.7950E-02\\
     13 & -0.2751E+00 &  0.3766E+00 &  0.2644E-02 &     31 &  0.8286E+00 &  0.3208E+00 &  0.4973E-02\\
     14 & -0.2171E+00 &  0.4603E+00 &  0.5976E-02 &     32 &  0.8864E+00 &  0.2511E+00 &  0.2959E-02\\
     15 & -0.1496E+00 &  0.4742E+00 &  0.3926E-02 &     33 &  0.9561E+00 &  0.1883E+00 &  0.2852E-02\\
     16 & -0.9083E-01 &  0.4394E+00 &  0.4400E-02 &     34 &  0.1019E+01 &  0.8369E-01 &  0.1274E-02\\
     17 & -0.2869E-01 &  0.4533E+00 &  0.5302E-02 &     35 &  0.1067E+01 &  0.1186E+00 &  0.1113E-02\\
     18 &  0.3169E-01 &  0.5440E+00 &  0.6711E-02 &     36 &  0.1463E+01 &  0.8369E+00 &  0.1582E-01\\      
&&\\
\tablecomments{The same as Table~(\ref{pdftable}), except for the redshift bin with
$\langle z \rangle = 3.99$ (see Table~\ref{zbins}).} 
\end{tabular} 
\end{table*}

\end{document}